\documentstyle[12pt,aaspp4,epsf]{article}
\def\ocen{$\omega$ Centauri}
\def\fun#1#2{\lower3.6pt\vbox{\baselineskip0pt\lineskip.9pt
  \ialign{$\mathsurround=0pt#1\hfil##\hfil$\crcr#2\crcr\sim\crcr}}}
\def\lap{\mathrel{\mathpalette\fun <}}
\def\gap{\mathrel{\mathpalette\fun >}}

\slugcomment{Submitted to The Astronomical Journal}

\begin{document}

\title{The Stellar Dynamics of \ocen}

\author{David Merritt}
\affil{Department of Physics and Astronomy, Rutgers University,
    New Brunswick, NJ 08855}
\author{Georges Meylan}
\affil{European Southern Observatory, Karl-Schwarzschild-Stra{\ss}e 2, 
D-85748 Garching bei M\"{u}nchen, Germany}
\author{Michel Mayor}
\affil{Observatoire de Gen\`{e}ve, CH-1290 Sauverny, Switzerland}

\bigskip
\centerline{Rutgers Astrophysics Preprint Series No. 205}
\bigskip

\begin{abstract}
The stellar dynamics of \ocen~ are inferred from the 
radial velocities of 469 stars measured with CORAVEL 
(\cite{may97}).
Rather than fit the data to a family of models, we generate 
estimates of all dynamical functions nonparametrically, by 
direct operation on the data.
The cluster is assumed to be oblate and edge-on but mass is not 
assumed to follow light.
The mean motions are consistent with axisymmetry but the rotation is not 
cylindrical.
The peak rotational velocity is $7.9$ km s$^{-1}$ at $\sim 11$ 
pc from the center.
The apparent rotation of \ocen~ is attributable in part 
to its proper motion.
We reconstruct the stellar velocity ellipsoid as a function of position,
assuming isotropy in the meridional plane. 
We find no significant evidence for a difference between the 
velocity dispersions parallel and perpendicular to the meridional 
plane.
The mass distribution inferred from the kinematics is slightly more 
extended than, though not strongly inconsistent with, the luminosity 
distribution.
We also derive the two-integral distribution function $f(E,L_z)$ 
implied by the velocity data.

\end{abstract}

\section{Introduction}

Since the mid-1960's, the standard method for analyzing 
globular cluster data has been a ``model-building'' approach
(\cite{kin66}; \cite{gun79}).
One begins by postulating a functional form for the phase-space 
density $f$ and the gravitational potential $\Phi$;
often the two are linked via Poisson's equation,
i.e. the stars described by $f$ are assumed to contain all of the 
mass that contributes to $\Phi$.
This $f$ is then projected into observable space and its 
predictions compared with the data.
If the discrepancies are significant, the model is rejected and 
another one is tried.
If no combination of functions $\{f,\Phi\}$ from the adopted family can be 
found that reproduce the data, one typically adds extra degrees of 
freedom until the fit is satisfactory.
For instance, $f$ may be allowed to depend on a larger number of 
orbital integrals (\cite{lup87}) or the range of 
possible potentials may be increased by postulating additional 
populations of unseen stars (\cite{dac76}).

This approach has enjoyed considerable popularity, in part
because it is computationally straightforward but also because 
(as \cite{kin81} has emphasized) globular cluster data are generally 
well fit by the standard models.
But one never knows which of the assumptions underlying the 
models are adhered to by the real system and which are not.
For instance, a deviation between the surface density profile of 
a globular cluster and the profile predicted by an isotropic 
model is sometimes taken as evidence that the real cluster is 
anisotropic.
But it is equally possible that the adopted form for $f(E)$ is 
simply in error, since by adjusting the dependence of $f$ on 
$E$ one can reproduce any density profile without anisotropy.
Even including the additional constraint of a measured velocity dispersion 
profile does not greatly improve matters since it is always 
possible to trade off the mass distribution with the velocity anisotropy in 
such a way as to leave the observed dispersions unchanged 
(\cite{dej92}).
Conclusions drawn from the model-building studies are hence very 
difficult to interpret; they are valid only to the extent that 
the assumed functional forms for $f$ and $\Phi$ are correct.

These arguments suggest that it might be profitable to interpret 
kinematical data from globular clusters in an entirely different 
manner, placing much stronger demands on the data and making fewer 
ad hoc assumptions about $f$ and $\Phi$.
Ideally, the unknown functions should be generated 
nonparametrically from the data.
Such an approach has rarely been tried in the past because of
the inherent instability of the deprojection process.
But techniques for dealing with ill-conditioned inverse problems 
are now well developed (e.g. \cite{wah90}; \cite{sco92}; \cite{gre94}), 
and the only thing standing in the way of a 
fully nonparametric recovery of $f$ and $\Phi$ is the limited 
size of most kinematical data sets.

As we show here, the $469$ stellar velocities measured by CORAVEL in 
\ocen~ (\cite{may97}, Paper I) permit the dynamical inverse problem for 
this system to be solved with relatively few ad hoc assumptions.
\ocen~ is a particularly good candidate for such a study 
since the two-body relaxation time $t_R$ is 
of order $10^9$ years even in its core, and $t_R$ exceeds a Hubble 
time at the half-light radius (e.g. \cite{mey95}).
Thus there is little justification for assuming that the dynamical 
state of \ocen~ is well described by a model derived from the theory of 
stellar encounters.
Instead, we expect the structure of this cluster to reflect in large measure 
the details of its formation process, about which very little is known.
The inverse-problem approach is also well suited to \ocen~ since 
the CORAVEL data extend, more or less uniformly, 
over the entire image of the cluster (see Figure 1).

Our analysis will proceed by discrete steps, each of 
which takes us deeper into the dynamics though at the price of a 
progressively larger set of assumptions.
As a first step we recover the rotational velocity field, 
assuming only that \ocen~ is axisymmetric and edge-on.
(The assumption that we lie in the equatorial plane of \ocen~ 
is made throughout this paper, partly for reasons of mathematical 
convenience, but also because \ocen~ is among the flattest of 
globular clusters and hence it is unlikely that we are viewing 
it from a direction far out of the equatorial plane.)
We next recover the two components $\sigma_{\varpi}$ and 
$\sigma_{\phi}$ of the stellar velocity ellipsoid,
after imposing dynamical equilibrium and assuming isotropy in the 
meridional plane, i.e. $\sigma_{\varpi}=\sigma_z$.
The latter assumption is the least restrictive one we can 
make which still allows us to uniquely recover the internal 
velocity dispersions from the observed ones.
The two-dimensional gravitational potential and the mass density then follow 
without additional assumptions; we emphasize that these are bona-fide 
dynamical estimates and are independent of any assumptions about the 
mass-to-light ratio or the stellar mass function.
Finally, we recover $f$ by assuming that the phase space density 
depends only on the orbital energy $E$ and angular momentum $L_z$ -- 
consistent with our previous assumption that $\sigma_{\varpi}=\sigma_z$, 
but slightly more restrictive in the sense that even three-integral $f$'s 
can be isotropic in the meridional plane.
At no point do we write down ad hoc forms for any of the unknown 
functions; the only restriction we impose is smoothness,
enforced via penalty functions.
Details of the technique may be found below and in Merritt (1996).

One advantage of analyzing the data in this way is that the 
conclusions drawn at every step are much easier to interpret than 
if $f$ and $\Phi$ were written down at the outset.
For instance, the dependence of the mean streaming velocity 
$\overline{v}_{\phi}$ on position in the meridional plane 
follows immediately from the velocity data
without any assumptions other than symmetry, 
and the uncertainties in our recovery of this function are due 
almost entirely to finite-sample fluctuations (which we estimate via the 
bootstrap) -- there are no biases like those that would result 
from the adoption of a rigid functional form for 
$\overline{v}_{\phi}$ or $f$. 
In the same way, our conclusions about the degree of velocity 
anisotropy in \ocen~ are based on direct reconstruction of the 
velocity ellipsoid, and not merely on deviations of the density 
profile from the predictions of an ad hoc isotropic model.

By proceeding in this step-by-step way, we also preserve the 
important distinction between the ``knowability'' of different 
dynamical quantities.
For instance, the phase-space density $f$ is related to the data by 
what is effectively a high-order differentiation, which means 
that recovery of $f$ from kinematical data will always be 
extremely ill-conditioned (\cite{mil63}).
It is therefore proper to save the determination of $f$ until the 
last, after quantities that are more directly related to the 
data -- like the rotational velocity field -- have been derived.
This distinction between well-determined and poorly-determined 
quantities is lost in the model-building approach, which begins 
by specifying $f$ and then treats all quantities that are derived 
from it on an equal footing.

We find that \ocen~ can be well described as an ``isotropic oblate 
rotator,'' i.e. a system in which the velocity residuals about 
the mean motions are approximately isotropic.
Our conclusion is stronger than one derived from the tensor 
virial theorem alone (\cite{mey86}), since we are able to make 
statements about the detailed dependence of the velocity 
ellipsoid on position.
We find in addition that the mass distribution in \ocen~ is 
consistent with the luminosity distribution of the bright stars, 
and that the mass density normalization is consistent with that 
predicted by the virial theorem under the assumption that mass 
follows light.
Hence there is no compelling reason to postulate a population of 
unseen objects in this cluster.
Finally, we show that the velocity data are fully consistent with 
a two-integral stellar distribution function $f(E,L_z)$.

In \S2 we review the CORAVEL data, which are described more fully 
in Paper I, and briefly describe the penalized likelihood 
formalism that is used in the reconstruction of the dynamical 
quantities.
The stellar density profile is derived in \S3; the rotational velocity 
field in \S4; and the velocity dispersions in \S5. 
Dynamical estimates of the gravitational potential 
and the mass distribution are presented in \S6, and the stellar 
distribution function is derived in \S7.
\S8 sums up and discusses prospects for future work.

\section{Data and Method}

The data on which this study is based are presented in Paper I.
The photoelectric spectrometer CORAVEL (\cite{may85}) was 
used to obtain 1701 radial velocity measurements of 483 
bright stars, mostly giants and subgiants, of which 469 were 
later determined to be members.
The typical measurement uncertainty is $\lap 1.0$ km s$^{-1}$.
The positions of the member stars are shown in Figure 1.
Previous analyses of various subsets of these data have been 
presented by Meylan \& Mayor (1986), Meylan (1987) and Meylan et 
al. (1995).

We adopt Cartesian coordinates $(X,Z)$ on the plane of the sky;
$Z$ is coincident with the isophotal minor axis of \ocen, as 
defined below, and $X$ increases westward.
The coordinate parallel to the line of sight is then $Y$.
We take minutes of arc as our units for $X$ and $Z$.
Adopting a heliocentric distance to \ocen~ of 5.2 kpc 
(\cite{mey87}), we find $1' = 1.51$pc.
The canonical ``core radius'' of \ocen~ is $\sim3'\approx4$ pc 
(\cite{pet75}) (although we show below that the brightness profile
of \ocen~ is consistent with a luminosity density that increases 
monotonically toward the center).
The tidal radius $r_t$ is roughly $50'\approx 75$ pc (\cite{pet75}).
For comparison, the velocity data in Paper I extend to about 
$1/2 r_t$, or $\sim 25'$ (Figure 1).

We assume throughout most of this study that \ocen~ is axisymmetric.
Cylindrical coordinates in the meridional plane are $(\varpi, 
z)$; the $z$ axis is parallel to the $Z$ axis, i.e. coincident with 
the isophotal minor axis.

Our treatment of the data will follow the nonparametric approach 
developed in an earlier series of papers (\cite{mer93b}, 1996; 
\cite{mer94}).
One seeks a smooth function -- call it $g$ -- such that the 
projection of $g$ into observable space is consistent with the data.
For instance, $g$ might be the rotational velocity field in the 
meridional plane, to be determined from a set of measured
positions and radial velocities.
The standard technique for solving such inverse problems
is to vary $g$ so as to minimize a functional like
\begin{equation}
-\log{\cal L}_p = \sum_{\rm data}\left[ g_p-{\bf A}g\over\epsilon\right]^2 + 
\alpha P(g),
\end{equation}
the ``penalized log likelihood'' (\cite{tho90}).
Here ${\bf A}$ represents the projection operator that brings $g$ 
into observable space; $g_p$ is the observed quantity; $\epsilon$ 
is a measurement error; and $P$ is a function that 
assigns a large penalty to noisy solutions.
The penalty function is needed since direct deprojection of 
the data is generally ill-conditioned, leading to physically 
useless solutions (\cite{kin81}).
In the implementations presented here, $P$ will depend on $g$ via its 
mean square second derivatives, a fairly standard choice (e.g. 
\cite{wah80}).
The degree of smoothness of the solution is then controlled by 
adjusting $\alpha$. 
One generally chooses $\alpha$ to be as small as possible 
consistent with smoothness so as to avoid biasing the solution.

Implicit in this approach is the assumption that the observable 
quantities $g_p$ contain sufficient information to uniquely constrain 
the unknown functions $g$.
The requisite proofs for the axisymmetric inverse problem 
are presented in Merritt (1996).

\section{Luminosity Density Profile}

The first inverse problem to be solved is the derivation of the 
spatial luminosity density in \ocen, $\nu(\varpi,z)$, as a function of 
position in the meridional plane, given measurements of the surface 
brightness $\Sigma(X,Z)$ on the plane of the sky.
Ideally we would adopt for $\Sigma$ the density of 
stars that comprise our kinematical sample, computed via a 
penalized-likelihood scheme (\cite{mer94}).
However the velocity sample was not specifically chosen to be 
magnitude-limited (see Paper I), nor is it as large as one 
would like for the nonparametric reconstruction of $\nu$.
Instead we follow the practice of Meylan (1987) and use
surface brightness determinations from a number of other sources.
These include centered aperture photometry (\cite{gas56}; 
\cite{dac79}) and drift scan measurements (\cite{dac79}) 
in the central regions, and star counts (\cite{kin68}) in 
the outer regions.
The normalized surface brightness measurements and their 
estimated errors are presented in Table 1 of Meylan (1987)
at 44 distinct radii $R_i$.
The studies from which these measurements were taken generally 
ignored the noncircular shape of \ocen, and so the radii
tabulated by Meylan (1987) and used here should be viewed as 
approximate averages over position angle.

\ocen~ is however significantly flattened.
Geyer, Hopp \& Nelles (1983) measured the isophotal ellipticity 
$\epsilon=1-b/a$ as a function of radius between $1.4'$ and 
$28'$ via photographic photometry.
They found a mean ellipticity of $0.121$; however $\epsilon$ 
varies significantly with radius, from $\epsilon\approx 0$ within 
$2'$ to $\epsilon\approx 0.25$ at $10'$, becoming rounder again 
at large radii.
The position angle of the principal axis was not found to vary 
significantly with radius.
Geyer et al. do not quote a value for this angle, but 
White \& Shawl (1987) derive an orientation of 6$^{\circ}$ 
east-from-north for the isophotal minor axis, and this value appears to be 
consistent with Figure 1 from Geyer et al.
We will adopt the White \& Shawl orientation in what follows.

Ideally, one would carry out a full deprojection 
of the surface brightness data $\Sigma_i$, measured at 
some set of points $\{X_i,Z_i\}$, to obtain an estimate of the 
space density $\nu(\varpi,z)$.
This inverse problem has a formally unique solution for any axisymmetric 
system that is viewed edge-on, even if its cross-section is not 
elliptical (\cite{ryb86}; \cite{ger96}).
However no two-dimensional surface brightness data for \ocen~ 
have been published.
Instead, we will assume that the density of stars in \ocen~ is 
stratified on similar, concentric oblate spheroids of axis ratio 
$b/a=0.879$, the average value found by Geyer et al. (1983) ---
consistent with our assumption that the equatorial plane is 
parallel to the line of sight.

The binned surface brightness measurements from Table 1 of Meylan (1987) 
are plotted in Figure 2a.
The solid line in that figure is an estimate of the surface 
brightness profile, defined as the function $\hat\Sigma(R)$ 
that minimizes the quantity
\begin{equation}
-\log{\cal L}_p = \sum_i 
\epsilon_i^{-2}\left[\log\Sigma_i-\log\Sigma(R_i)\right]^2 + 
\alpha\int \left[{d^2(\log\Sigma)\over d(\log 
R)^2}\right]^2 d\log R.
\label{opt1}
\end{equation}
Here $\epsilon_i$ is the estimated uncertainty in the
surface brightness measurement $\log\Sigma_i$ at $R_i$.
The value of $\alpha$ used to derive the profile of
Figure 2a was selected via the generalized cross-validation technique 
(\cite{wah90}, ch. 4); this value minimizes the estimated, integrated 
square error in $\hat\Sigma(R)$.

The estimate of $\nu(r)$ may be defined as the Abel inversion of 
the estimate $\hat\Sigma(R)$:
\begin{equation}
\hat\nu(r) = -{1\over\pi}\int_r^{\infty} {d\hat\Sigma\over dR} 
{dR\over\sqrt{R^2-r^2}}.
\end{equation}
Here $r$ is also an azimuthally-averaged mean radius.
But deprojection has the property of amplifying the noise in the 
data, and we expect that the optimal value of $\alpha$ to be 
applied to the surface brightness data when deriving $\nu$
will be larger than the optimal value for estimating $\Sigma$ itself
(e.g. \cite{sco92}, p. 132).
Unfortunately, there is no generally-accepted criterion for 
selecting $\alpha$ in cases like this (\cite{wah90}, p. 105).
To produce the estimate of $\nu(r)$ shown in Figure 2b, we used a 
value of $\alpha$ roughly three times the optimal value for estimation
of $\Sigma(R)$.
The 95\% confidence bands in that figure were derived via the 
bootstrap (\cite{wah90}, p. 71), with a pointwise correction for 
bias (as in \cite{sco92}, p. 259).

The stellar density profile $\nu(r)$ is evidently very well determined by 
these data at all radii outside of $\sim 1'$.
(For comparison, \cite{pet75} find $\sim3'$ for the \ocen~ core radius.) 
However at smaller radii the density profile is poorly determined, due 
primarily to the amplification of uncertainties resulting from 
the deprojection.
The profile of Figure 2b actually has a power-law cusp, $\nu\sim 
r^{-1}$, inside of $0.5'$; however the confidence bands are 
consistent with a wide range of slopes in this region, including 
even a profile that declines toward the center.

Henceforth we take as our estimate of the stellar number density 
$\nu(\varpi,z)=\hat\nu(\xi)$, with $\xi^2=(1-\epsilon)\left[\varpi^2 + 
z^2/(1-\epsilon)^2\right]$ and $\hat\nu$ the function of Figure 
2b.
Almost all of the stars with measured velocities lie outside of 
the region where the uncertainties in $\nu$ are significant; 
furthermore, the uncertainties in most of the quantities derived 
below will be affected much more by limitations in the 
kinematical sample than by errors in $\nu$.
Hence we will ignore uncertainties in $\nu$ in what follows.

\section{Rotational Velocity Field}

Next we wish to find the dependence of the mean azimuthal 
velocity $\overline{v}_{\phi}$ on position in the meridional plane.
As a first step, we investigate the variation of the mean 
line-of-sight velocity $\overline{V}_Y(X,Z)$ over the image of \ocen.
An estimate of $\overline{V}_Y$ can be 
defined as the function $\hat{\overline{V}}_Y$ that minimizes
\begin{eqnarray}
-\log{\cal L}_p & = & \sum_i
\epsilon_i^{-2}\left[V_i-\overline{V}_Y(X_i,Z_i)\right]^2 \nonumber \\ 
& + & \alpha\int\int\left[\left({d^2\overline{V}_Y\over dX^2}\right)^2
+ 2\left({d^2\overline{V}_Y\over dX dZ}\right)^2
+ \left({d^2\overline{V}_Y\over dZ^2}\right)^2\right] dX dZ.
\label{opt2}
\end{eqnarray}
Here $\epsilon_i$ is the estimated uncertainty in the stellar velocity 
$V_i$ measured at point $(X_i, Z_i)$.
The function $\hat{\overline{V}}_Y$ that minimizes 
(\ref{opt2}) is a so-called ``thin-plate smoothing spline'' 
(\cite{wah90}, p. 30).
Figure 3a shows the result; the smoothing parameter $\alpha$ was 
again chosen via generalized cross-validation.
The rotational velocity field is approximately symmetric but 
shows what appears to be a systematic twist at large radii.

Here we note a correction that must be made to the measured velocities
before proceeding further.
\ocen~ has a substantial proper motion.
Feast, Thackeray \& Wesselink (1961) note that the translation 
of a solid object produces an apparent rotation, since the 
projection of the space velocity along the line of sight is 
different at different points in the image.
This ``perspective rotation'' results in a radial velocity increment
\begin{equation}
V_{pr} = {X\over D} V_X + {Z\over D} V_Z
\label{prot}
\end{equation} 
where $V_X$ and $V_Z$ are the components of the cluster space velocity 
parallel to the $X$ and $Z$ axes and $D$ is the distance to the 
cluster.
Curves of constant $V_{pr}$ are straight lines of slope 
$-V_X/V_Z$, and the apparent ``rotation axis'' is oriented at an angle 
$\theta_{pr} = \tan^{-1}(-V_X/V_Z)$ measured clockwise from the $Z$-axis.
As a result of the perspective rotation, the cluster appears to 
rotate as a solid body, with the rotational velocity increasing 
linearly with radius.

The proper motion of \ocen~ has been determined by Murray, Jones 
\& Candy (1965).
Corrected for differential Galactic rotation (Cudworth 1994),
the components of the proper motion parallel to our $X$ and $Z$ axes are
\begin{equation}
\mu_X = 0.26 \pm 0.06\ {\rm arcsec}/{\rm century},\ \ 
\mu_Z = -0.74 \pm 0.05\ {\rm arcsec}/{\rm century}.
\end{equation}
Using our adopted distance of 5.2 kpc and assuming an error in 
this number of $\pm$ 10\%, we find
\begin{equation}
V_X = 64.1 \pm 16\ {\rm km\ s}^{-1},\ \ \ \ V_Z = -183.0 \pm 22\ {\rm km\ 
s}^{-1}.
\end{equation}
The resulting change in the radial velocity at point $(X,Z)$ is
\begin{equation}
\left( 0.0187 \pm 0.005\ X' - 0.053 \pm 0.006\ Z'\right) {\rm km\ 
s}^{-1}
\label{pmxz}
\end{equation}
with $X'$ and $Z'$ measured in minutes of arc from the cluster 
center.
The amplitude of the effect is small but not negligible, 
exceeding 1 km s$^{-1}$ near the edge of our kinematical sample.
The spurious ``rotation axis'' is oriented at 19 degrees 
west-from-north, or about 25 degrees in a clockwise direction 
from the isophotal minor axis.
At points near the $X$ axis, the sign of the induced radial velocity 
is opposite to that of the true, line-of-sight rotational velocity.
Thus the perspective rotation would be expected to induce a 
slight twist at large radii in the contours of constant mean 
radial velocity --- and in just the direction observed in Figure 3a.

Figure 4a shows the corrected rotational velocity field, 
computed by minimizing (\ref{opt2}) after removing the 
contribution of the perspective rotation to the measured velocities.
The contours are now substantially more symmetric at large radii.
The rotational velocity along the isophotal major axis, shown in 
Figure 4c, has a peak value of about 7 km s$^{-1}$ at a radius of 
$7'\approx 11$ pc; this profile is reasonably symmetric about 
the cluster minor axis.

If \ocen~ were exactly axisymmetric, the kinematic minor axis 
would coincide with the isophotal minor axis.
Figure 4a shows that this is approximately true, although there 
still appears to be a separation of a few degrees between the axis of 
maximal apparent rotation and the isophotal major axis (the $X$-axis
in Figure 4).
The discrepancy is probably not significant given the relatively small 
number of velocities that determine our estimate of the rotational velocity
field, and given the likely error in White \& Shawl's (1987) 
determination of the minor axis.
Furthermore the proper motion adopted here might be in error, leading
to an incorrect adjustment for the perspective rotation.
In any case, we will continue to use the White \& Shawl 
estimate of the minor axis orientation in what follows.

Our next task is to recover $\overline{v}_{\phi}(\varpi,z)$, the mean 
rotational velocity in the meridional plane.
This inverse problem has a formally unique solution if \ocen~ is 
axially symmetric and edge-on (\cite{mer91}).
However because $\overline{v}_{\phi}$ is related to the data via 
a deprojection, it will be intrinsically less well determined 
than $\overline{V}_Y$.

We seek the function $\hat{\overline{v}}_{\phi}(\varpi,z)$ that minimizes
\begin{equation}
-\log{\cal L}_p=\sum_i \epsilon_i^{-2}\left(V_i - {\cal L}_i {\overline 
v_{\phi}}\right)^2 +
\alpha\int\int\left[\left({\partial^2{\overline
v}_{\phi}\over \partial\varpi^2}\right)^2 + 
2\left({\partial^2{\overline
v}_{\phi}\over \partial\varpi\partial z}\right)^2 +
\left({\partial^2{\overline v}_{\phi}\over \partial 
z^2}\right)^2\right]
d\varpi dz,
\label{opt3}
\end{equation}
where ${\cal L}$ is the line-of-sight projection operator
\begin{equation}
{\cal L}_i{\overline v_{\phi}} =
2X_i\Sigma(X_i,Z_i)^{-1}\int_{X_i}^{\infty}\nu(\varpi,Z_i){\overline{v}}_{\phi
}(\varpi,Z_i) {d\varpi\over\sqrt{\varpi^2-X_i^2}}.
\label{proj3}
\end{equation}
Equation (\ref{proj3}) assumes that the cluster is axially 
symmetric, and that the observer lies in the equatorial plane.
We can decrease the noise in the estimate by assuming in addition 
that the rotational velocity field is symmetric about the equatorial plane,
an assumption that makes dynamical sense and that is consistent with
the appearance of the contours in Figure 4.
One way to enforce this constraint is to reflect all of the data
into one quadrant of the $(X,Z)$ plane, 
with appropriate changes in the sign of $V_i$.
This procedure was found to give reasonable results, except that 
the estimate $\hat{\overline v}_{\phi}$ so produced had a 
discontinuous gradient along the $\varpi$ and $z$ axes.
Physically more reasonable results were obtained by changing only 
the signs of the $X_i$, and requiring (via a set of linear constraints) 
that the solution be symmetric about $z=0$ and zero along the $z$ axis.
The penalty function was then able to enforce nearly-zero 
$z$-derivatives on the equatorial plane, as desired.

The corresponding, constrained optimization problem was solved
via quadratic programming (\cite{mer96}).
Various values were tried for the smoothing parameter; 
Figure 5 shows the result for one choice of $\alpha$ judged 
nearly optimal.
The contours of constant $\hat{\overline{v}}_{\phi}$ are 
remarkably similar in 
shape to those of the parametric model postulated by Meylan \& 
Mayor (1986), at least in the region near the center where the 
solution is strongly constrained by the data.
As noted by them, the rotational velocity field is clearly not cylindrical;
instead, $\overline{v}_{\phi}$ has a peak value of 
$\sim 8$ km s$^{-1}$ at about $7'$ from the center in the 
equatorial plane, and falls off both with increasing $\varpi$ and $z$.
In the region inside the peak, the rotation is approximately solid-body.

One would like to estimate the uncertainties associated with 
the solution presented in Figure 5.
The error in $\hat{\overline{v}}_{\phi}$ can be broken into two 
parts: the ``variance,'' i.e. the mean square fluctuation 
resulting from the finite sample size and measurement errors; 
and the ``bias,'' the 
systematic deviation of the solution from the true 
$\overline{v}_{\phi}$ (e.g. \cite{mul88}, p. 29).
The variance is possible to estimate via the 
bootstrap, by generating a large number of pseudo data sets from 
the smooth solution and observing how greatly the estimates of 
$\overline{v}_{\phi}$ fluctuate from sample to sample.
The bias is a tougher nut to crack.
While the techniques used here are (unlike parametric methods) 
unbiased in the limit of large sample sizes, 
for finite samples the smoothing will introduce a systematic 
deviation of the solution from the true $\overline{v}_{\phi}$.
The situation is complicated still more by the fact that the 
amplitude of the variance depends (inversely) on the amount of 
smoothing, and hence on the bias.
Thus a too-large choice for $\alpha$ will artificially reduce the 
variance that would otherwise be expected from a sample of a 
given size.
(Exactly the same is true for a histogram -- the $\sqrt{N}$ uncertainty 
associated with any bin depends on the bin width, i.e. on the degree 
of smoothing.)

We nevertheless constructed formal confidence bands
on our estimate of $\overline{v}_{\phi}$
according to the following scheme (\cite{wah90}, p. 71).
We emphasize that intervals constructed in this way are 
confidence intervals for the {\it estimate} $\hat{\overline{v}}_{\phi}$ 
and not for the true function $\overline{v}_{\phi}$, and 
do not reflect the error due to the bias.
We expect the latter to be greatest at large radii, where the 
data constrain the solution the least.

\noindent
1. Compute a smooth estimate of $\overline{v}_{\phi}$; in our 
case, this estimate is the function plotted in Figure 5.

\noindent
2. Project this estimate into observable space and record 
the predicted, mean velocity $\hat{\overline{V}}_i$ at the 
positions $(X_i,Z_i)$ of the observed stars.

\noindent
3. Generate a large number $N$ of pseudo data sets 
$\tilde{V}_i=\hat{\overline{V}}_i + \tilde{\epsilon}_i$, where 
$\tilde{\epsilon}_i$ is a random number from the normal 
distribution ${\cal N}(0,\hat{\sigma}^2_i)$, and $\hat{\sigma}^2_i$ 
is an estimate of the variance of the measured velocities about 
their mean value at $(X_i,Z_i)$.
This variance was set equal to the square of the estimated 
velocity dispersion displayed in Figure 4b.
(Measurement errors in the $V_i$ are of order 1 km s$^{-1}$ and
are negligible compared to the intrinsic velocity dispersion.)

\noindent
4. For each of these pseudo data sets, compute a new estimate of 
$\overline{v}_{\phi}$ and record the values at each point on the 
solution grid.

\noindent
5. The $\delta$ confidence intervals at any 
point on the grid are given by the $\delta/2\ N$th and 
$(1-\delta/2)\ N$th values from the distribution of estimates 
$\hat{\overline{v}}_{\phi}$ at that point.

\noindent
6. Since the pseudo data are generated from the smoothed estimate 
$\hat{\overline{v}}_{\phi}$, which is biased, the estimates 
generated from the pseudo data will be ``doubly'' biased.
One can compute the bias between the original estimate and the 
bootstrap estimates and then subtract it from the computed 
confidence bands (e.g. \cite{sco92}, p. 259).
This (slightly ad hoc) procedure yields confidence bands that are 
correctly situated with respect to the original estimate, rather 
than with respect to the average of the bootstrap estimates.

Figure 6 gives 95\% confidence intervals so constructed on 
$\hat{\overline {v}}_{\phi}$, along two lines in the meridional plane:
as a function of $\varpi$ at $z=0$, and as a function of $z$ 
at the value of $\varpi$ corresponding to peak rotation in the 
equatorial plane.
The confidence intervals on $\hat{\overline{v}}_{\phi}$ are wide, 
partly a consequence of the fact that relatively little of the 
motion in \ocen~ is ordered.
The peak rotational velocity is $7.9_{-2.5}^{+2.4}$ km s$^{-1}$ $(95\%)$.
For comparison, Meylan \& Mayor (1986), using an ad hoc 
parametric form for $\overline{v}_{\phi}$, derived a peak value of 
$\sim 8$ km s$^{-1}$ at $\sim 7'$, in excellent agreement with our result.
Outside $\sim 15'$, the confidence bands shown in Figure 6 suggest 
that the form of the rotational velocity field is only weakly 
constrained by these data.

We would like to compare our inferred rotational velocity field 
in \ocen~ to the predictions of a theoretical model.
However it is not clear that any relevant models exist.
Because of the long relaxation time in \ocen, the rotation
probably still reflects to a large extent the state of the cluster 
shortly after its formation.
Our estimate of $\hat{\overline{v}}_{\phi}(\varpi,z)$ might therefore be most 
useful as a constraint on cluster formation models.

A simpler question is whether \ocen~ can be described as an 
``isotropic oblate rotator''; that is, whether the distribution 
of velocity residuals about the mean motion are approximately the 
same in all directions.
We will answer this question in the affirmative below, by 
showing that there is no evidence for a significant difference 
between the velocity dispersions parallel and perpendicular to 
the meridional plane.
Hence \ocen~ can reasonably be described as ``rotationally 
flattened.''
Meylan \& Mayor (1986) reached a similar conclusion after noting 
a strong correlation of the local rotational velocity with 
the isophotal flattening.

\section{Velocity Dispersions}
We now wish to dig deeper into the dynamics 
and reconstruct the dependence of the 
velocity dispersions on position in the meridional plane.
We again start by investigating the variation in the plane of the 
sky.
We seek the function $\hat{\overline{V_Y^2}}(X,Z)$ that minimizes
\begin{eqnarray}
-\log{\cal L}_p & = & \sum_i
\epsilon_i^{-4}\left[V_i^2-\overline{V_Y^2}(X_i,Z_i)\right]^2 \nonumber \\ 
& + & \alpha\int\int\left[\left({d^2\overline{V_Y^2}\over dX^2}\right)^2
+ 2\left({d^2\overline{V_Y^2}\over dX dZ}\right)^2
+ \left({d^2\overline{V_Y^2}\over dZ^2}\right)^2\right] dX dZ;
\label{opt4}
\end{eqnarray}
then $\hat{\sigma}_Y^2=\hat{\overline{V_Y^2}} - 
{\hat{\overline{V}}_Y}^2$.
The result is shown in Figures 3b and 4b, before and after 
removal of the perspective rotation.
The contours of constant $\hat\sigma_Y$ are reasonably symmetric, 
at least near the center; however there is no clear indication of an 
elongation of the contours in the $X$-direction, parallel to the 
isophotal major axis.
The velocity dispersion along the major axis 
(Figure 4d) falls from a maximum of $\sim 17$ km s$^{-1}$ at the 
center to $\sim 8$ km s$^{-1}$ at $15'$.

Meylan et al. (1995, Figure 1) present a circularly-symmetrized 
velocity dispersion profile derived from the same data used here.
They note that the velocity dispersion rises sharply very near 
the center of \ocen, from $\sim 17$ km s$^{-1}$ at $2'$ to $\sim 
22$ km s$^{-1}$ within $1'$.
This central value is derived from only 16 stars and has a formal 
($1\sigma$) uncertainty of $\pm 3.9$ km s$^{-1}$; thus the 
increase near the center may not be significant.
We can mimic the effect of a fine radial grid like the one used 
in Meylan et al. (1995) by reducing the value 
of the smoothing parameter $\alpha$.
We find that our estimate of the central dispersion rises as 
$\alpha$ is reduced, and we can easily reproduce the Meylan et al. 
value for $\sigma_Y(0)$ if $\alpha$ is chosen to be sufficiently small.
However the velocity dispersion profile so produced is very 
noisy and almost certainly undersmoothed; the high central 
dispersion has the appearance of a finite-sample fluctuation.
We conclude that there is no secure case to be made from 
the current data for a sharply-rising velocity dispersion within 
the inner minute of arc.

The line-of-sight velocity dispersions 
contain contributions from both $\sigma_{\varpi}$ and 
$\overline{v_{\phi}^2}$; $\sigma_z$ does not contribute due to our 
assumption that the observer sits in the equatorial 
plane.
If we assume in addition that $\sigma_{\varpi}=\sigma_z\equiv\sigma(\varpi,z)$, 
i.e. that the velocity ellipsoid is circular in the meridional 
plane, it becomes possible to infer both 
$\sigma(\varpi,z)$ and $\sigma_{\phi}(\varpi,z)$ independently from 
$\overline{V_Y^2}(X,Z)$ (\cite{mer96}).
The assumption of isotropy in the meridional plane is physically 
restrictive and one would like to avoid making it.
However, allowing radial anisotropy adds {\it two} unknown functions to 
be determined from the data: $\sigma_z(\varpi,z)$ and 
$\overline{v_{\varpi}v_z}(\varpi,z)$, which together with $\sigma_{\varpi}$
and $\sigma_{\phi}$ define the 3-D shape and orientation of the 
velocity ellipsoid.
The dynamical inverse problem under this less-restrictive set of 
assumptions has not been formulated but one would certainly need 
much more data than are currently available in \ocen~ in order to 
find a unique solution.
This issue is discussed at greater length in \S 8.

Following the scheme in Merritt (1996), 
we search for the functions $\hat{\sigma}^2$ and
$\hat{\overline{v^2_{\phi}}}$ that minimize
\begin{eqnarray}
-\log{\cal L}_p & = & \sum_i \epsilon_i^{-4}\left[{V}_i^2 - {\cal L}_i
\left\{ \sigma^2, \overline{v^2_{\phi}}\right\} \right]^2 
\nonumber \\
& + & \alpha\int\int\left[\left({\partial^2 \sigma^2\over 
\partial\varpi^2}
\right)^2 + 2\left({\partial^2 \sigma^2 \over 
\partial\varpi\partial z}
\right)^2 + \left({\partial^2 \sigma^2 \over \partial 
z^2}\right)^2\right]
d\varpi dz \nonumber \\
& + & 
\alpha\int\int\left[\left({\partial^2\overline{v^2_{\phi}}\over 
\partial
\varpi^2}\right)^2 + 2\left({\partial^2 
\overline{v^2_{\phi}}\over\partial\varpi
\partial z}\right)^2 + \left({\partial^2 \overline{v^2_{\phi}} 
\over \partial z^2}
\right)^2\right] d\varpi dz
\label{opt5}
\end{eqnarray}
subject to the constraints
\begin{equation}
{\partial\nu\over\partial\varpi}{\partial\sigma^2\over\partial z} 
-
{\partial\nu\over\partial z}{\partial\sigma^2\over\partial\varpi} 
+
{\nu\over\varpi}{\partial\over\partial z}\left(\sigma^2 -
\overline{v^2_{\phi}}\right) = 0
\label{magic}
\end{equation}
which are satisfied for any axisymmetric system with 
$\sigma_{\varpi}=\sigma_z\equiv\sigma$.
The projection operator ${\cal L}_i$ becomes 
\begin{equation}
{\cal L}_i\left\{\sigma^2,\overline{v^2_{\phi}}\right\} =
2\Sigma(X_i,Z_i)^{-1}\int_{X_i}^{\infty}\nu(\varpi,Z_i)\left[\left(
1-{X_i^2\over \varpi^2}\right)\sigma^2(\varpi,Z_i) +
{X_i^2\over\varpi^2}\overline{v^2_{\phi}}(\varpi,Z_i)\right]
{d\varpi\over\sqrt{\varpi^2-X_i^2}},
\label{projb}
\end{equation}
and $\hat\sigma_{\phi}^2=\hat{\overline{v_{\phi}^2}} - 
\hat{\overline v}_{\phi}^2$.
This is again a quadratic programming problem.
The results are shown in Figure 7.
The vertical gradients of $\hat{\sigma}$ and $\hat{\sigma}_{\phi}$ were 
forced to be zero along the $\varpi-$axis.
There are obvious differences between the contour shapes for 
$\hat{\sigma}$ and $\hat{\sigma}_{\phi}$, though both functions 
are reasonably symmetric.

We would again like to know which of the details of Figure 7 are 
securely implied by the data and which are due to finite-sample 
fluctuations.
Perhaps the most interesting question here is the value of the 
anisotropy parameter $\gamma=1-\sigma_{\phi}^2/\sigma^2$, the 
degree to which the azimuthal and meridional dispersions differ.
Figures 8 and 9 give estimates of $\sigma$, $\sigma_{\phi}$ and 
$\gamma$ on the major and minor axes, along with 95\% bootstrap 
confidence bands.
Although $\hat{\sigma}_{\phi}$ is less than $\hat{\sigma}$ at 
most radii, the confidence bands are wide and there is no clear 
indication of anisotropy anywhere in the meridional plane.
Near the center, $\sigma_{\phi}$ could be as small as 60\% of 
$\sigma$ or as large as 1.4$\sigma$ (95\%).
The formal 95\% confidence interval on $\sigma(0,0)$ 
is $17^{+2.1}_{-2.6}$ km s$^{-1}$.

The simplest interpretation of these results is that \ocen~ is an
``isotropic oblate rotator,'' i.e. that the velocity residuals 
about the mean motion are isotropically distributed everywhere.
However we emphasize the uncertainties associated with this 
interpretation.
We have ruled out radial anisotropies by fiat, and even under this 
restrictive assumption, the confidence bands on $\hat{\sigma}$ and 
$\hat{\sigma}_{\phi}$ are quite wide.
Clearly, a somewhat larger kinematical sample than ours would be 
needed to place reasonably strong constraints on the degree of anisotropy
in \ocen.

Meylan (1987) noted that anisotropic models from the
Michie-King family could be made to fit the surface brightness
data for \ocen~ only if the velocity ellipsoid was allowed to
become significantly elongated outside of $\sim 10'$.
Meylan's result highlights the limited range of models
from the Michie-King (or any similar) family.
In the absence of a priori information about the likely
functional form of $f(E,L)$, the surface brightness profile
of a stellar system does not contain any useful information
about the degree of velocity anisotropy.
Particularly in clusters like \ocen, where the collisional 
relaxation
time is long, conclusions derived from comparison with
collisional models should probably not be given much weight.

\section{Mass Distribution}

Having obtained estimates of the first and second velocity 
moments of the stellar distribution function,
we can now construct dynamical estimates of the gravitational potential 
$\Phi(\varpi,z)$ and the mass density $\rho(\varpi,z)$.
The former follows most directly from the Jeans equation relating the vertical 
gradients in the stellar pressure to the vertical force:
\begin{equation}
\nu{\partial\Phi\over\partial 
z}=-{\partial(\nu\sigma^2)\over\partial z},
\end{equation}
or
\begin{equation}
\hat{\Phi}(\varpi,z) = \int_z^{\infty}\hat{\sigma}^2(\varpi,z)
{\partial\log\hat{\nu}\over \partial z}dz - \hat{\sigma}^2(\varpi,z).
\end{equation}
We have again made use of our assumption that 
$\sigma_{\varpi}=\sigma_z\equiv\sigma$.
The mass density follows from Poisson's equation,
\begin{equation}
\hat{\rho}(\varpi,z) = {1\over 4\pi
G}\left[{1\over\varpi}{\partial\over\partial\varpi}
\left(\varpi{\partial\hat{\Phi}\over\partial\varpi}\right) + 
{\partial^2\hat{\Phi}\over
\partial z^2}\right].
\end{equation}
Estimates of $\Phi$ and $\rho$ so derived are bona-fide dynamical 
estimates; they are independent of any assumptions 
about the relative distributions of mass and light, the stellar 
mass function, etc.
However these estimates {\it are} dependent on our neglect of 
anisotropy in the meridional plane, a point that we return to below.

Figure 10 gives the bias-corrected estimates of $\Phi$ and 
$\rho$.
Also shown in Figure 10 are the potential and mass distribution that 
\ocen~ would have if $\rho$ were proportional to $\nu$, i.e. if mass 
followed light.
The total mass of \ocen~ under the latter assumption follows immediately
from the virial theorem (Appendix; \cite{mey86}): 
it is $1.45\times 10^4 \langle V_Y^2\rangle M_{\odot}\approx 
2.86\times 10^6 M_{\odot}$, with $\langle V_Y^2\rangle$ the mean 
square line-of-sight velocity expressed in ${\rm km}^2/{\rm s}^2$.
The dynamically-inferred potential is slightly more elongated
at large radii, $r\gap 5'$, than the ``potential'' generated 
by the light.
The differences between $\hat{\rho}(\varpi,z)$ and $\hat{\nu}(\varpi,z)$ 
are somewhat greater, due in part to the fact that $\hat{\rho}$ is a second 
derivative of $\hat{\Phi}$.
The dependence of both functions on radius along the major axis 
and their confidence bands are shown in Figure 11.
The dynamically inferred mass distribution is less centrally concentrated 
than the light distribution.
However this difference is hardly significant given 
the width of the confidence bands on $\hat\rho$.
Furthermore, we would expect the smoothing to reduce the degree 
of central concentration in the inferred mass, and our ``bias 
correction'' probably does not account fully for this systematic 
error.
Thus we are inclined to interpret Figure 11 
conservatively, i.e. to conclude that there is no evidence for a 
significant difference between the mass and light distributions 
in \ocen.

In any case, our formal, 95\% confidence interval on the central mass 
density of \ocen~ is $2110^{+530}_{-510}M_{\odot}{\rm pc}^{-3}$.
This is consistent with an estimate based on the 
``core-fitting'' formula, $\rho(0)\approx 9\sigma_Y^2(0)/4\pi G 
R_c^2$ (\cite{kin66}).
Adopting $R_c=3.0'=4.6$ pc and $\sigma_Y=17$ km s$^{-1}$, 
we find $\rho(0)= 2270 M_{\odot} {\rm pc}^{-3}$.
This agreement is to be expected since the assumptions underlying 
the core-fitting formula -- velocity anisotropy and a constant 
$M/L$ -- are consistent with our analysis of the \ocen~ data.

Although we have failed to falsify the mass-follows-light model 
for \ocen, we stress that very different mass distributions are
certain to be consistent with our limited data.
By far the greatest uncertainty in estimates of $\Phi$ and $\rho$ 
for hot stellar systems results from lack of knowledge about the 
radial elongation of the velocity ellipsoid.
In a spherical system, allowing $\sigma_r/\sigma_{\phi}$ to have 
any physically permissible value leads to order-of-magnitude uncertainties 
in the total mass or the central density as derived from the velocity 
dispersion profile (\cite{dej92}).
The same is likely to be true in axisymmetric systems, with the 
role of the radial anisotropy played by the elongation of 
the velocity ellipsoid in the meridional plane.
If the velocity ellipsoid became radially elongated at large 
radii, for instance, the inferred mass density profile would be 
less centrally concentrated 
than what we found assuming $\sigma_{\varpi}=\sigma_z$.
Possible routes for overcoming this indeterminacy are discussed in
\S 8.

\section{Distribution Function}

Our final step is to estimate the formally unique, two-integral 
distribution function $f(E,L_z)$ that generates the inferred 
stellar number density $\hat{\nu}(\varpi,z)$ and rotational velocity 
field $\hat{\overline{v}}_{\phi}(\varpi,z)$ in the inferred potential 
$\hat{\Phi}(\varpi,z)$.
We are justified in expressing $f$ as a function of only two 
integrals since any $f(E,L_z)$ implies a velocity ellipsoid that 
is isotropic in the meridional plane, consistent with our 
assumption that $\sigma_{\varpi}=\sigma_z$.
However there also exist three-integral $f's$ with this property 
and so the distribution function that we derive here is not 
strictly unique even given our restrictive assumption about the 
kinematics.
Our goal is simply to show that a non-negative, two-integral 
$f$ exists that is consistent with our estimates of $\nu$, 
$\overline{v}_{\phi}$ and $\Phi$.
The precise functional form of $f(E,L_z)$ is of less interest, for two reasons. 
First, we have already derived the lowest-order moments of $f$, 
i.e. $\nu, \overline{v}_{\phi}, \sigma$ and $\sigma_{\phi}$.
Thus $f$ itself tells us relatively little that we do not already 
know.
Second, like the mass density $\rho$, $f$ is related to the data 
via a high-order differentiation and hence its derivation is 
extremely ill-conditioned.
We expect that many different expressions for $f(E,L_z)$ -- obtained 
with different smoothing parameters, or different penalty 
functions, etc. -- will be almost equivalent in terms of their 
ability to reproduce the input functions $\hat\nu$ and 
$\hat{\overline{v}}_{\phi}$ and in fact we found this to be the case.

Figure 12 nevertheless shows estimates of the even and odd parts of $f$:
\begin{equation}
f(E,L_z)=f_+(E,L_z) + f_-(E,L_z), \ \ \ \ f_{\pm}(E,L_z)\equiv 
{1\over 2}
\left[f(E,L_z)\pm f(E,-L_z)\right],
\end{equation}
the former derived from $\hat{\nu}$ and the latter from 
$\hat{\overline{v}}_{\phi}$ (\cite{mer96}).
These estimates were constrained to be everywhere positive.
The positivity constraint did not preclude the inferred $f's$ 
from reproducing the input functions with high accuracy, and we 
infer from this that \ocen~ can in fact be well described by a two-integral 
$f$.

\section{Conclusions and Directions for Future Work}

Our principle conclusions are summarized here.

1. The rotational velocity field in \ocen~ is consistent with 
axisymmetry, once a correction is made for ``perspective 
rotation'' resulting from the cluster's proper motion.
The rotation is strongly non-cylindrical, with a peak rotation 
speed of $7.9^{+2.4}_{-2.5}$ km s$^{-1}$ (95\%) at a distance 
of $\sim 11$ pc 
from the cluster center in the equatorial plane.
The rotation is approximately solid-body at small radii; at large 
radii, the available data do not strongly constrain the 
form of the rotational velocity field.

2. By assuming that the residual velocities are isotropic in the 
meridional plane, $\sigma_{\varpi}=\sigma_z\equiv\sigma$, we 
derived the dependence of the two independent velocity 
dispersions $\sigma$ and $\sigma_{\phi}$ on position in the 
meridional plane.
The central velocity dispersion parallel to the meridional plane 
is $\sigma(0,0)=17^{+2.1}_{-2.6}$ km s$^{-1}$.
There is no evidence for significant anisotropy anywhere in 
\ocen.
Thus, this cluster can reasonably be described as an ``isotropic oblate 
rotator.''

3. The gravitational potential and mass distribution in \ocen~ are 
consistent with the predictions of a model in which the mass is 
distributed in the same way as the bright stars.
The central mass density is $2110^{+530}_{-510}M_{\odot}{\rm 
pc}^{-3}$.
However this result may be strongly dependent on our assumption that the 
velocity ellipsoid is isotropic in the meridional plane.

4. We derive a two-integral distribution function $f(E,L_z)$ for the 
stars in \ocen~ and show that it is fully consistent with our 
data. 

Our method is based on the smallest set of assumptions that 
permit formally unique estimates of the first and 
second moments of the stellar velocity distribution in an 
edge-on, axisymmetric cluster, given observed values of
these moments on the plane of the sky.
The deprojection of the line-of-sight velocity dispersions can be 
carried out only if the velocity dispersions in the meridional 
plane are reducible to a single function of 
position $\sigma(\varpi,z)$, rather than the three functions 
$\{\sigma_{\varpi}, \sigma_z, \overline{v_{\varpi}v_z}\}$ that 
characterize a fully general axisymmetric system.
While there are many possible choices for the relation 
between these three functions, setting $\overline{v_{\varpi}v_z}=0$ and 
$\sigma_{\varpi}=\sigma_z$ as we do is the only choice consistent with
a general two-integral distribution function $f(E,L_z)$.
Nevertheless, some of our results -- particularly the form of the 
mass distribution -- might be strongly dependent on our neglect 
of velocity anisotropy in the meridional plane.
Future work on \ocen~ should therefore be directed toward 
obtaining enough kinematical data to independently constrain the 
functions $\sigma_{\varpi}$, $\sigma_z$ and $\overline{v_{\varpi}v_z}$.

One route would be to measure velocity dispersions parallel to
the plane of the sky from internal proper motions. 
Proper motion velocity dispersions have in fact been measured 
in a few globular clusters (e.g. M13, \cite{cum79}; M3, \cite{cud79}).
In a spherical system, such data allow one to infer the velocity 
anisotropy as a function of radius (\cite{leo89}), and this technique has 
been used to constrain the mass distribution in M13 (\cite{leo92}).
The corresponding existence proof has yet to be carried out for 
axisymmetric systems, but we note that proper motion velocity 
dispersions would contribute two new functions of two variables 
$\{\sigma_X(X,Z), \sigma_Z(X,Z)\}$, which might 
provide just enough information to uniquely constrain the two unknown 
functions $\sigma_z(\varpi,z)$ and $\overline{v_{\varpi}v_z}(\varpi,z)$.

Alternatively, one could try to obtain much larger samples of radial 
velocities in \ocen, of order $10^4$, which would permit the derivation of 
the full line-of-sight velocity distribution $N(X,Z,V_Y)$
at a discrete number of positions in the image of the cluster.
In fact the radial velocities of approximately $4200$ stars 
in the core of \ocen~ have already been measured 
using the Rutgers Fabry-P\'erot interferometer (\cite{geb97}),
and line-of-sight velocity distributions for these stars have been 
presented by Merritt (1997).
An observational program directed toward measuring a similar 
number of radial velocities at larger radii in \ocen~ is currently 
nearing completion (\cite{rei93}).
Here again, we lack a rigorous theoretical understanding of precisely what 
dynamical information are contained in the function $N(X,Z,V_Y)$.
However numerical experiments in the spherical geometry 
(\cite{mer93a}; \cite{ger93}) suggest that line-of-sight 
velocity distributions, like proper motions, contain 
considerable information about the velocity anisotropy.

The kinematical sample analyzed here is large by the standards 
of just a few years ago, yet still small enough that the confidence 
intervals on many quantities of interest are still quite wide.
We trust that this situation will be rectified in the very near 
future.

\bigskip\bigskip
We thank C. Pryor for comments on the manuscript.
This work was supported by NSF grant AST 90-16515 and NASA grant 
NAG 5-2803 to DM.

\clearpage

\centerline{\bf APPENDIX}

Here we derive an expression for the total mass of Omega Centauri 
assuming that the mass density is everywhere proportional to luminosity
density.

The scalar virial theorem is $2K+W=0$, with $K$ the total 
kinetic energy and $W$ the potential energy.
Since we have assumed that the starlight, and hence the mass, is 
stratified on similar oblate spheroids, $W$ can be written
\begin{equation}
W=-4\pi^2G {\cal S} \left({b\over a}\right){\sqrt{1-e^2}\over e}\sin^{-1}e
\end{equation}
(\cite{rob62})
where
\begin{equation}
{\cal S} = \int_0^{\infty} 
dm^2\rho(m^2)\int_0^{m^2}\rho(m'^2)m'dm'^2.
\label{sint}
\end{equation}
Here $\rho$ is the mass density, stratified on surfaces of constant 
$m^2=\varpi^2 + z^2/(1-e^2)$, and $e=\sqrt{1-b^2/a^2}$.
Using our adopted $b/a=0.879$ for Omega Centauri, 
and writing ${\cal S}=M^2 {\cal S}'$ with $M$ the total mass,
this becomes $W=-31.8 G M^2 {\cal S}'$.
We can evaluate ${\cal S}'$ using the number density profile 
derived in \S 3; the result is ${\cal S}'= 1.463\times 10^{-3}$ pc$^{-1}$, 
assuming a distance to Omega Centauri of 5.2 kpc.

The total kinetic energy is $K=2K_x+K_z$, where (as above) the $z$-axis  
is perpendicular to the equatorial plane.
The tensor virial theorem gives
\begin{equation}
K_z=\left({W_{zz}\over W_{xx}}\right) K_x
\end{equation}
with
\begin{equation}
{W_{zz}\over W_{xx}} = 2\left({b\over a}\right)^2 
{\left[{a\over b}-{\sin^{-1}e\over e}\right]\over 
\left[{\sin^{-1}e\over e} - {b\over a}\right]} = 0.9010
\end{equation}
(\cite{rob62}).
Thus $K=2.901 K_x= 1.451 M\langle V_Y^2\rangle$,
with $\langle V_Y^2\rangle$ the (mass weighted) mean square 
line-of-sight velocity of the stars.
Equating $2K$ and $|W|$ then gives
\begin{equation}
M=1.445\times 10^4 \langle V_Y^2\rangle M_{\odot},
\end{equation}
where $\langle V_Y^2\rangle$ is measured in km s$^{-1}$.
This estimate of the mass is independent of any assumptions about
the internal kinematics, though it is strongly dependent on the assumption
that mass follows light.

\clearpage

\figcaption[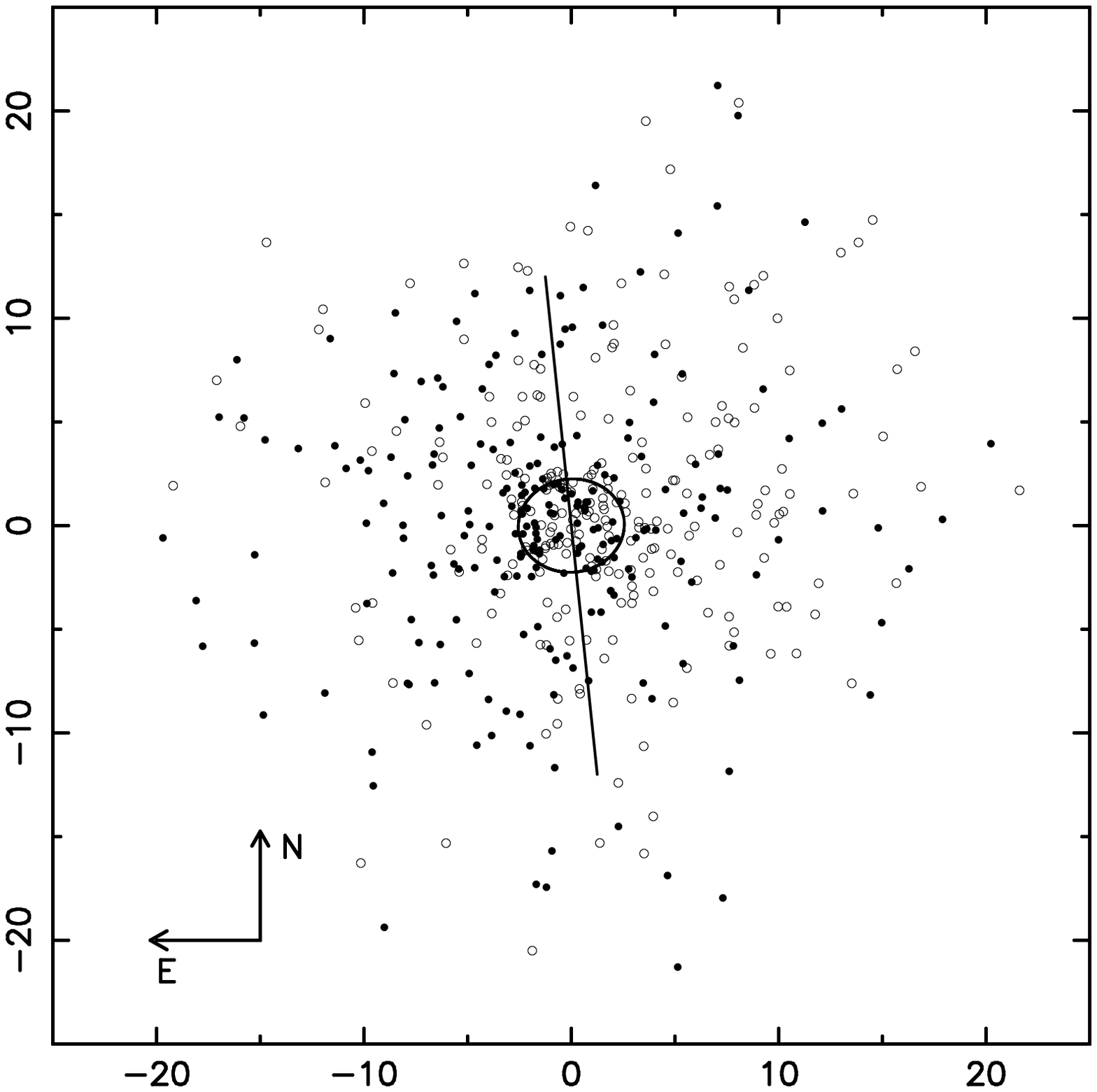]{\label{fig1}}Positions of stars with 
velocities measured by CORAVEL in \ocen.
Distances are measured in minutes of arc from the cluster center.
Filled/open circles are stars with radial velocities that are 
greater/less than the cluster mean.
The solid line shows the orientation of the photometric minor axis 
as determined by White \& Shawl (1987); this is the $Z$ axis in 
what follows.
The plotted ellipse has a mean radius equal to the \ocen~ core radius as 
determined by Peterson \& King (1975), and an eccentricity equal 
to the average value for \ocen, $\epsilon=0.121$, as determined 
by Geyer et al. (1983).

\figcaption[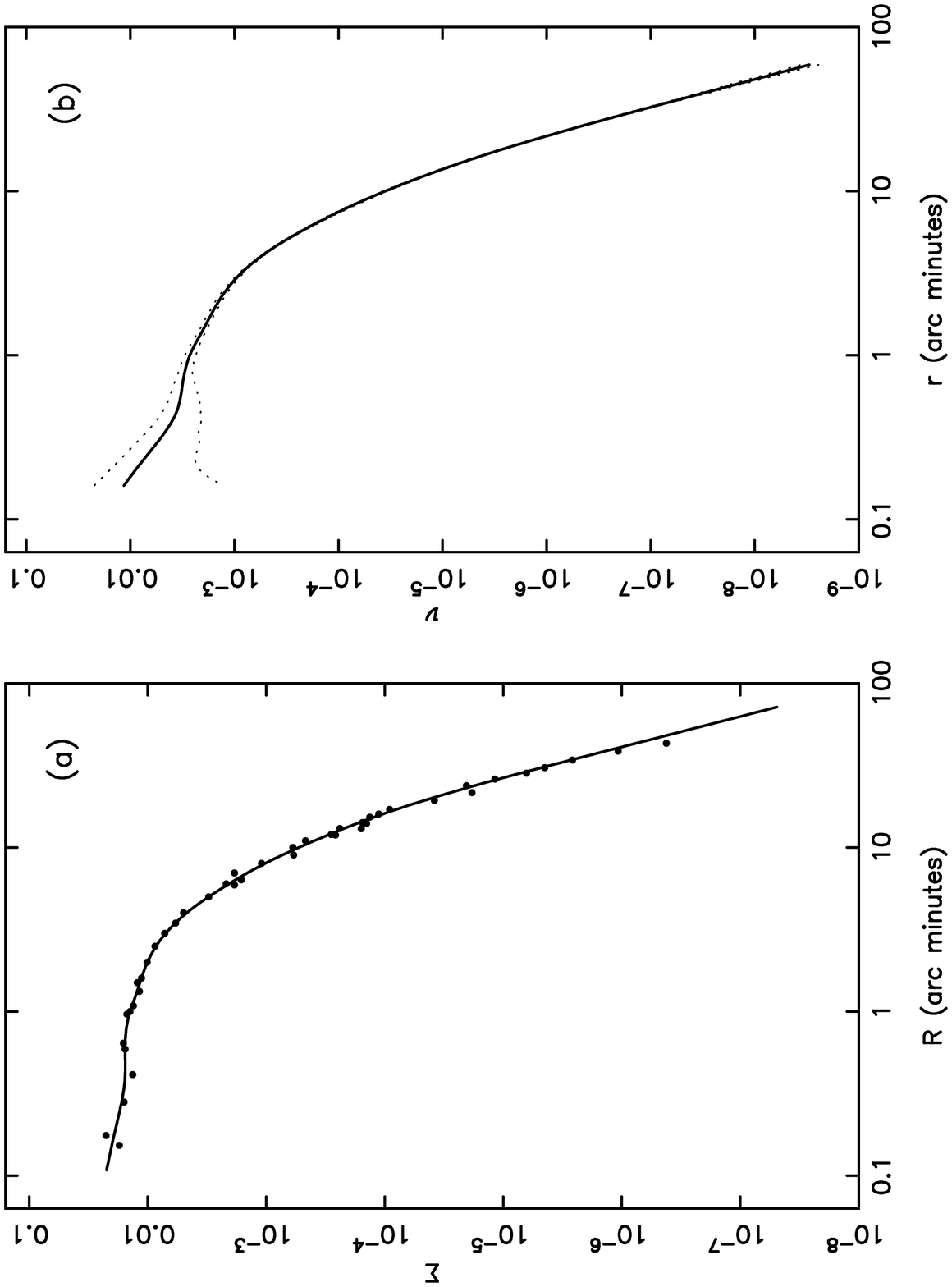]{\label{fig2}}Surface brightness (a) and space
density (b) profiles for \ocen.
The data in (a) are from Table 1 of Meylan (1987); the solid line is the 
solution to the optimization problem defined by Eq. (\ref{opt1}).
The dashed lines in (b) are 95\% confidence bands on the estimate 
of $\nu(r)$. 
Both profiles are normalized to unit total number.

\figcaption[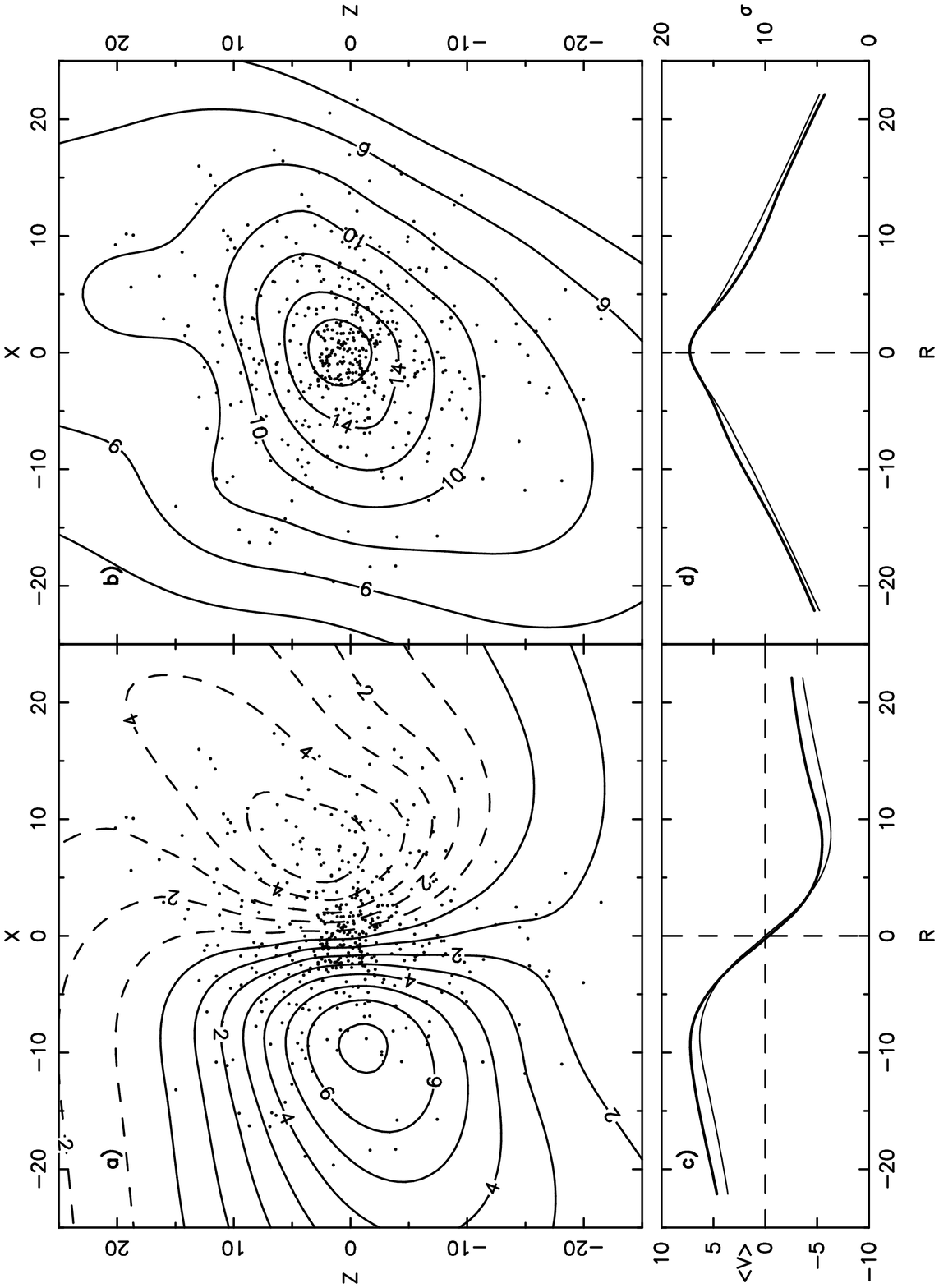]{\label{fig3}} (a) The line-of-sight, 
rotational velocity field of \ocen, obtained as the solution to 
the optimization problem (\ref{opt2}).
The $Z$-axis is parallel to the isophotal minor axis and east is 
toward the left; distances are measured in arc minutes.
Dots indicate positions of stars in the kinematical sample.
Contours are labelled in km s$^{-1}$, measured with respect to 
the cluster mean; dashed contours indicate negative velocities.
(b) The line-of-sight velocity dispersion field of \ocen, 
obtained as the solution to the optimization problem (\ref{opt5}).
Plotted are contours of the velocity dispersion about the local 
mean shown in (a); velocity errors have also been removed in 
quadrature.
(c) {\it Heavy line}: line-of-sight rotational velocity along the 
major axis of \ocen, in km s$^{-1}$; {\it thin line}: antisymmetrized profile.
(d) {\it Heavy line}: line-of-sight velocity dispersion along the 
major axis of \ocen; {\it thin line}: symmetrized profile.

\figcaption[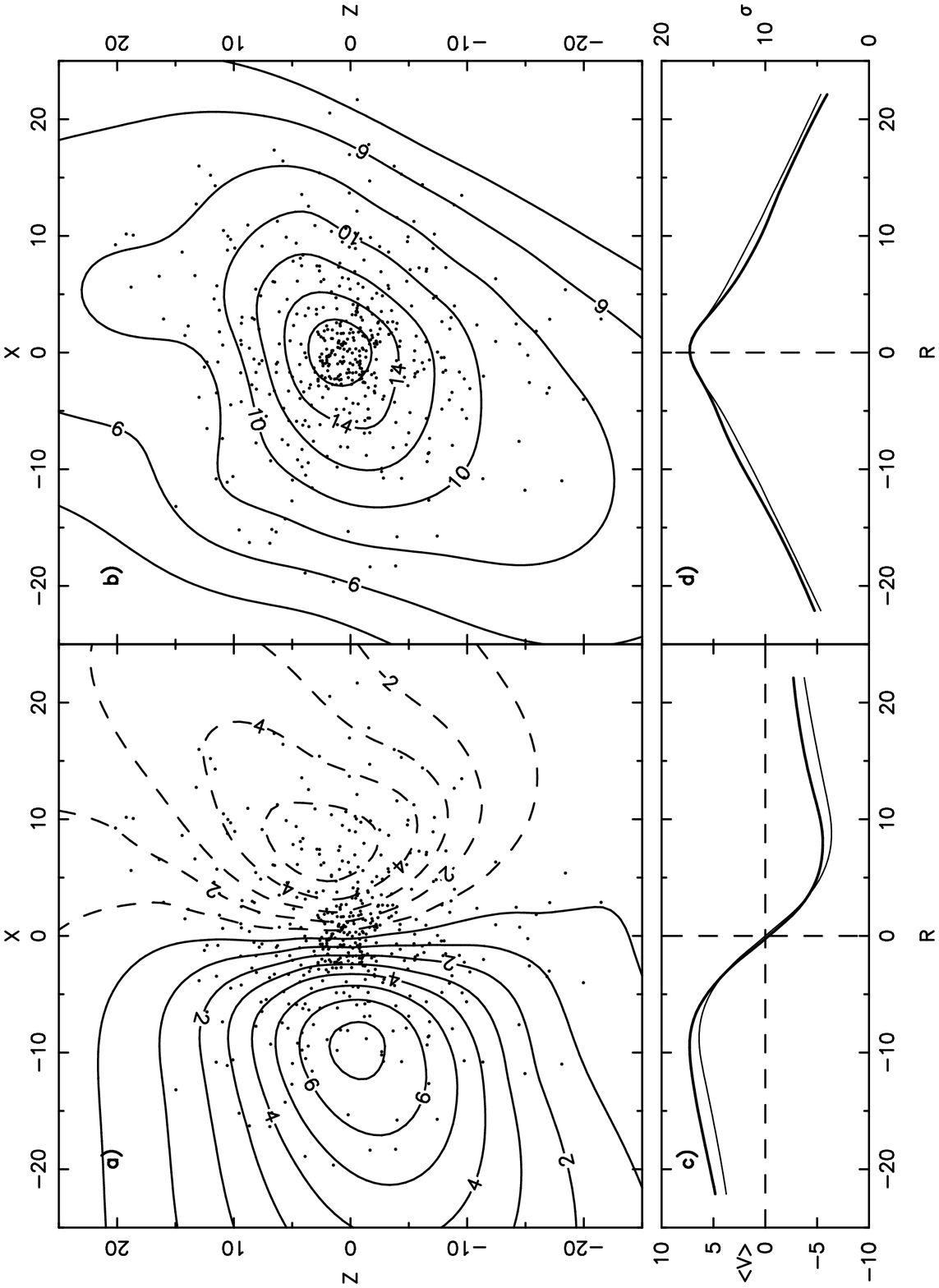]{\label{fig4}} Like Fig. 3, except that 
the perspective rotation due to the proper motion of \ocen~ has 
been removed from the measured velocities.

\figcaption[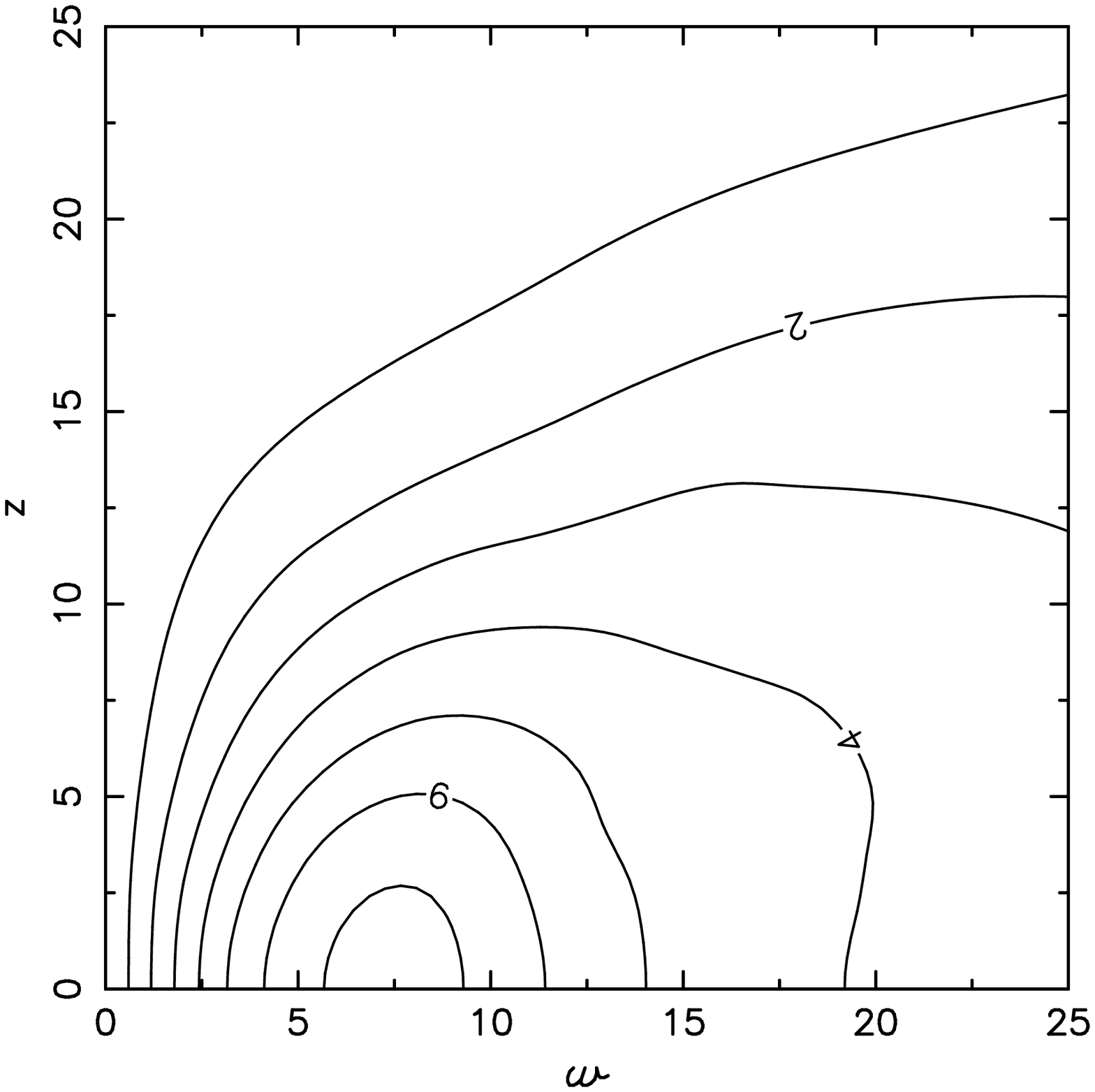]{\label{fig5}} 
Estimate of the mean azimuthal velocity $\overline{v}_{\phi}$ 
in the meridional plane of \ocen, obtained as the solution to the 
optimization problem (\ref{opt3}).
Distances are in arc minutes and contours are labelled in km 
s$^{-1}$.

\figcaption[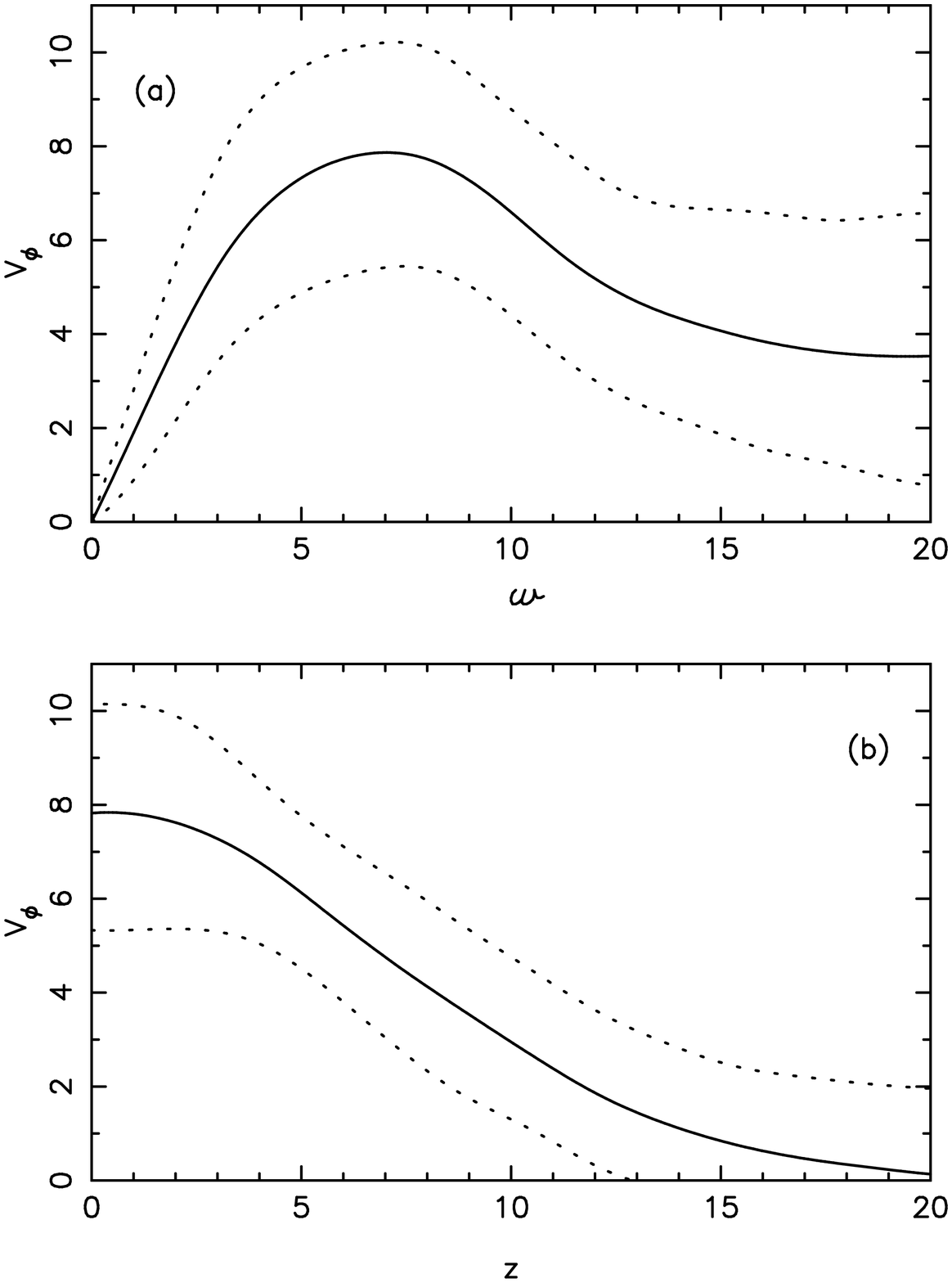]{\label{fig6}} (a) Mean azimuthal velocity 
in the equatorial plane.
b) Mean azimuthal velocity in the meridional plane 
as a function of $z$ at $\varpi=7'$.
Dotted lines are 95\% confidence bands, computed via the 
bootstrap.
Distances are expressed in minutes of arc.

\figcaption[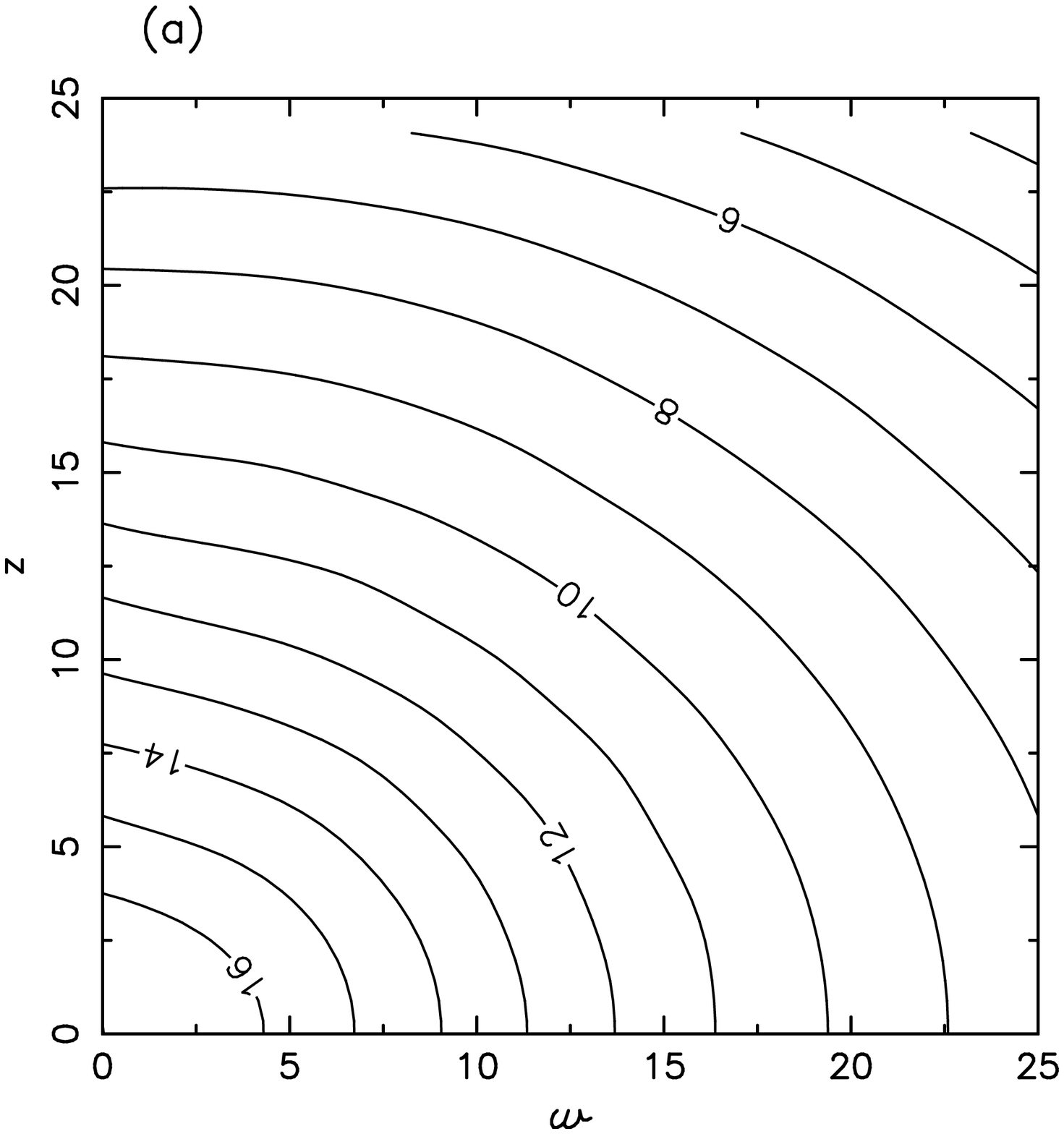]{\label{fig7}} 
Estimates of the meridional (a) and azimuthal (b) velocity dispersions, 
$\sigma$ and $\sigma_{\phi}$, in the meridional plane, 
obtained as solutions to the constrained optimization problem
defined by equations (\ref{opt5}) and (\ref{magic}).
Distances are expressed in minutes of arc and contours are labelled in km 
s$^{-1}$.

\figcaption[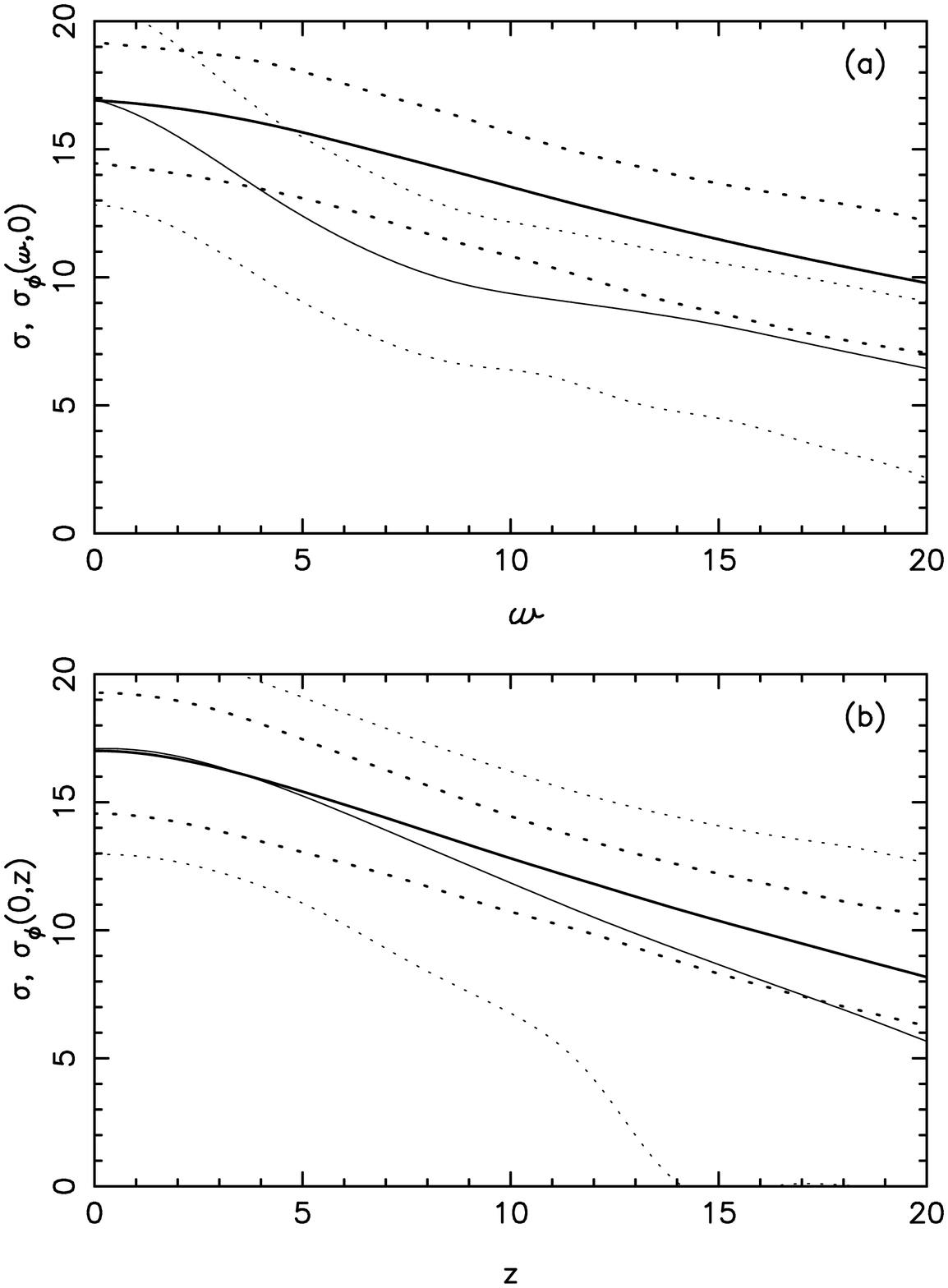]{\label{fig8}} Estimates of the velocity 
dispersion profiles along the major (a) and minor (b) axes in 
the meridional plane.
Thick lines: $\sigma$; thin lines: $\sigma_{\phi}$.
Dotted lines are 95\% confidence bands computed via the 
bootstrap.
Distances are expressed in minutes of arc.

\figcaption[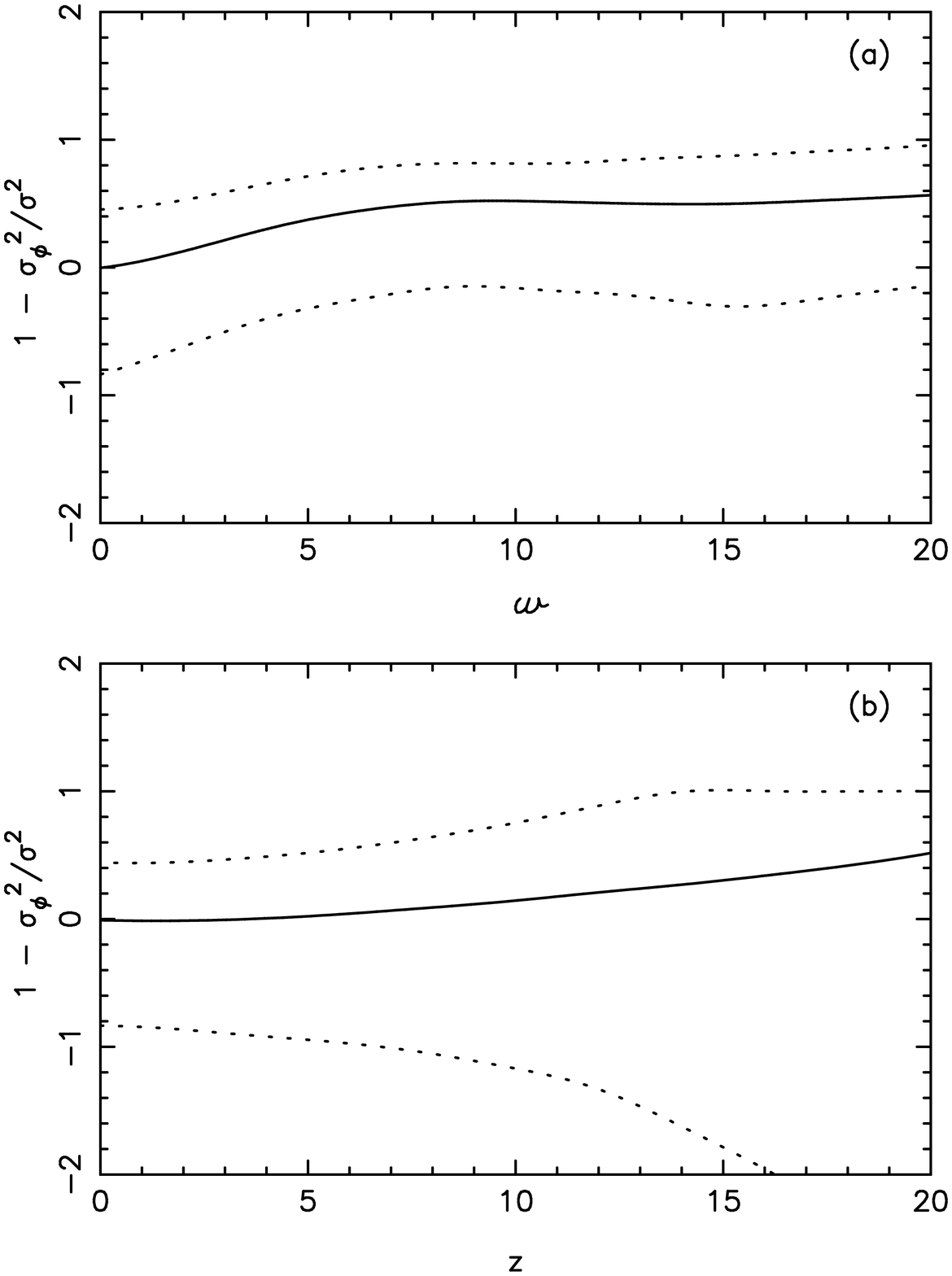]{\label{fig9}} Estimates of the velocity 
anisotropy $1-\sigma_{\phi}^2/\sigma^2$ along the major (a) and 
minor (b) axes in the meridional plane.
Dotted lines are 95\% confidence bands computed via the 
bootstrap.
Distances are expressed in minutes of arc.

\figcaption[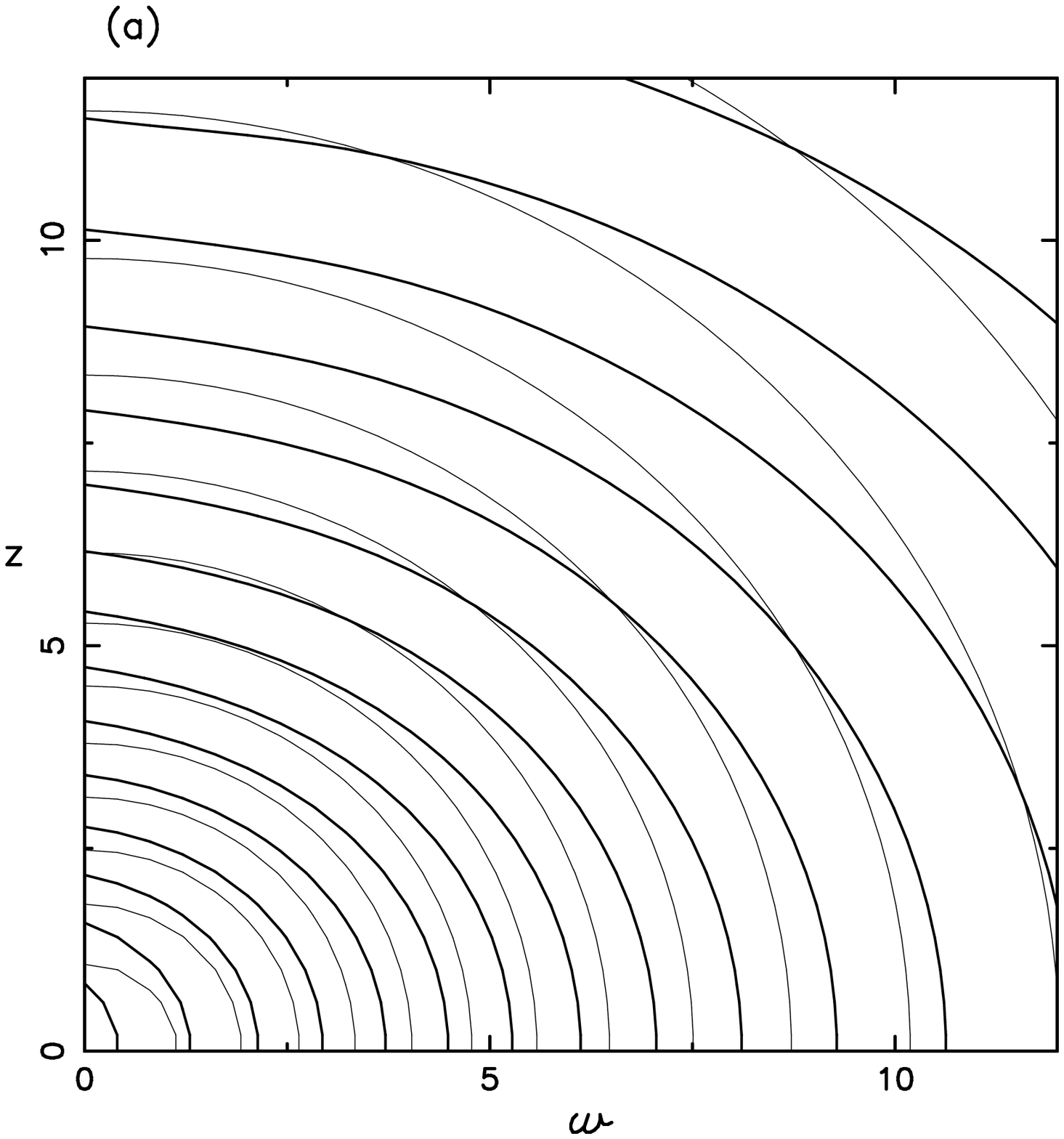]{\label{fig10}} Estimates of the gravitational
potential (a) and the mass density (b) derived from the Jeans and 
Poisson equations.
Thin lines are the corresponding functions derived under the assumption
that mass follows light in \ocen.
Distances are expressed in minutes of arc.

\figcaption[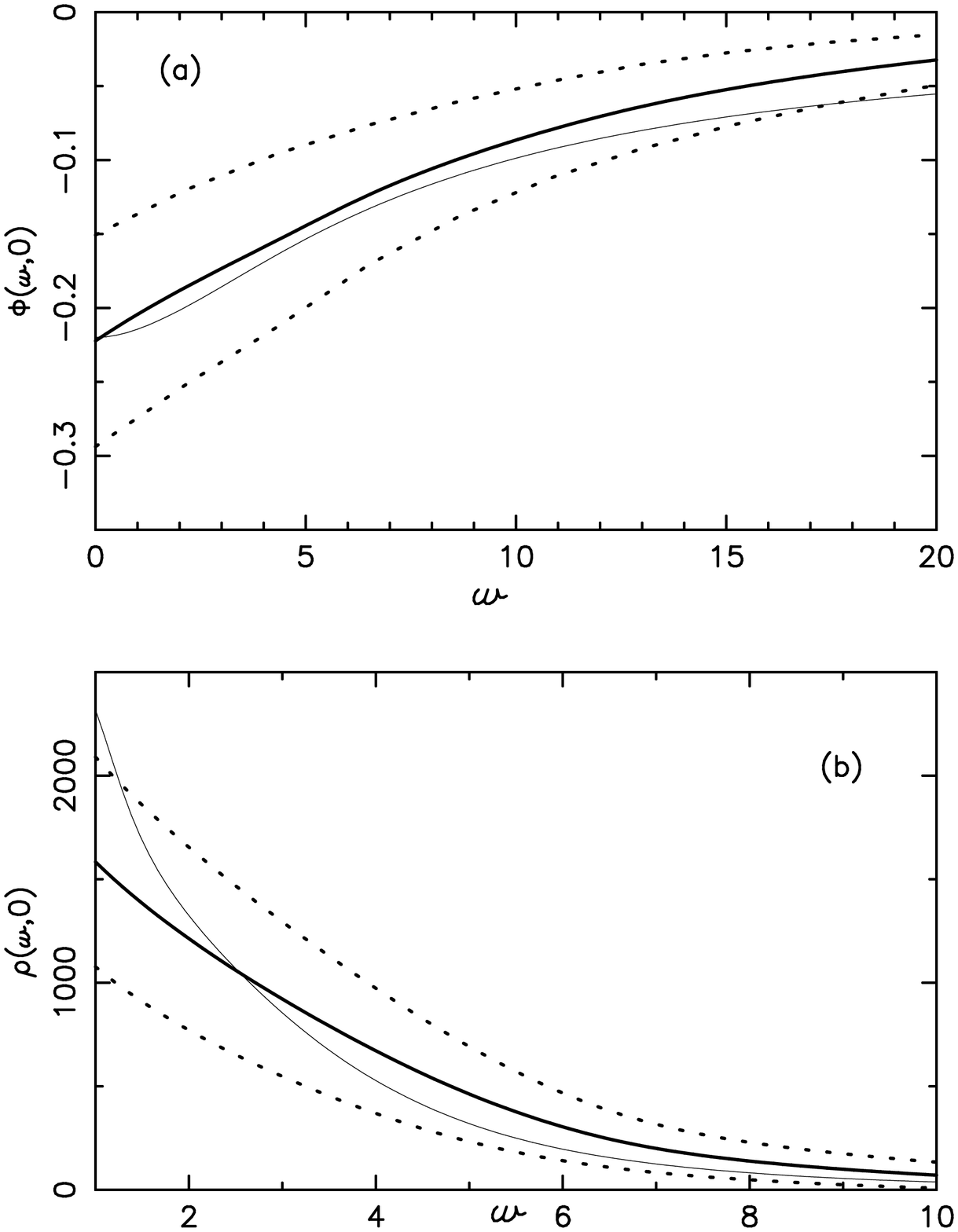]{\label{fig11}} Dependence of $\Phi$ and 
$\rho$ on distance along the major axis in \ocen.
Solid lines are the dynamically-derived estimates, with 95\% 
confidence bands (dashed).
Thin lines are the corresponding functions derived under the assumption
that mass follows light in \ocen.
Units of $\rho$ are $M_{\odot}pc^{-3}$; the potential is 
expressed in units such that the total mass of \ocen~ is unity.
Distances are expressed in minutes of arc.

\figcaption[figure12.ps]{\label{fig12}} Two-integral distribution 
functions for \ocen.
(a) $f_+(E,L_Z)$; (b) $f_-(E,L_z)$.
Heavy lines are the curves of maximum $L_z$ as a function of $E$.
Contours are separated by $0.27$ in $\log_{10} f$.
$E$ and $L_z$ are expressed in units such that the total mass of 
\ocen~ is unity.

\setcounter{figure}{0}

\begin{figure}
\plotone{figure1.ps}
\caption{ }
\end{figure}

\begin{figure}
\plotone{figure2.ps}
\caption{ }
\end{figure}

\begin{figure}
\plotone{figure3.ps}
\caption{ }
\end{figure}

\begin{figure}
\plotone{figure4.ps}
\caption{ }
\end{figure}

\begin{figure}
\plotone{figure5.ps}
\caption{ }
\end{figure}

\begin{figure}
\plotone{figure6.ps}
\caption{ }
\end{figure}

\begin{figure}
\plotone{figure7a.ps}
\caption{ }
\end{figure}

\setcounter{figure}{6}

\begin{figure}
\plotone{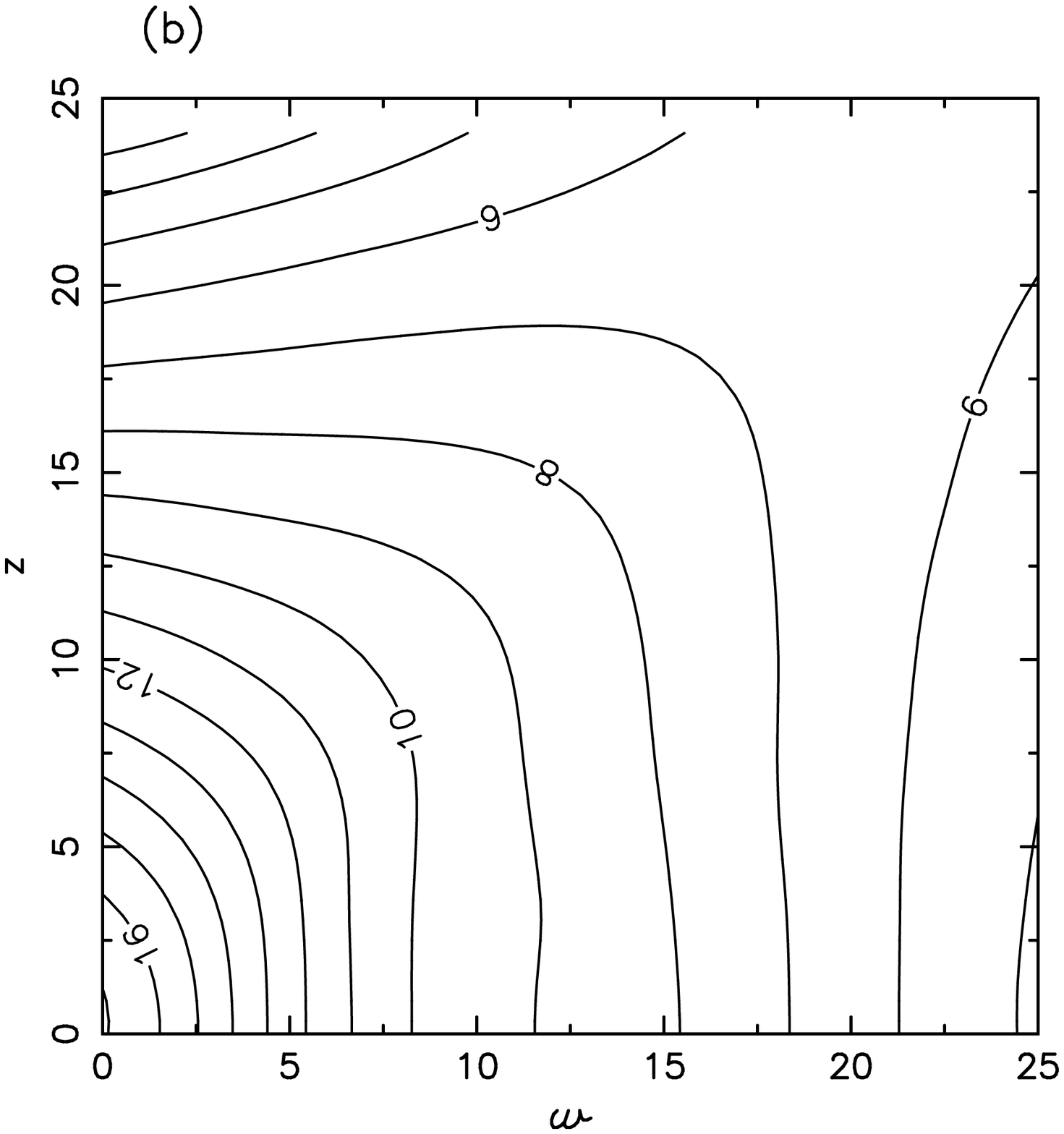}
\caption{ }
\end{figure}

\begin{figure}
\plotone{figure8.ps}
\caption{ }
\end{figure}

\begin{figure}
\plotone{figure9.ps}
\caption{ }
\end{figure}

\begin{figure}
\plotone{figure10a.ps}
\caption{ }
\end{figure}

\setcounter{figure}{9}

\begin{figure}
\plotone{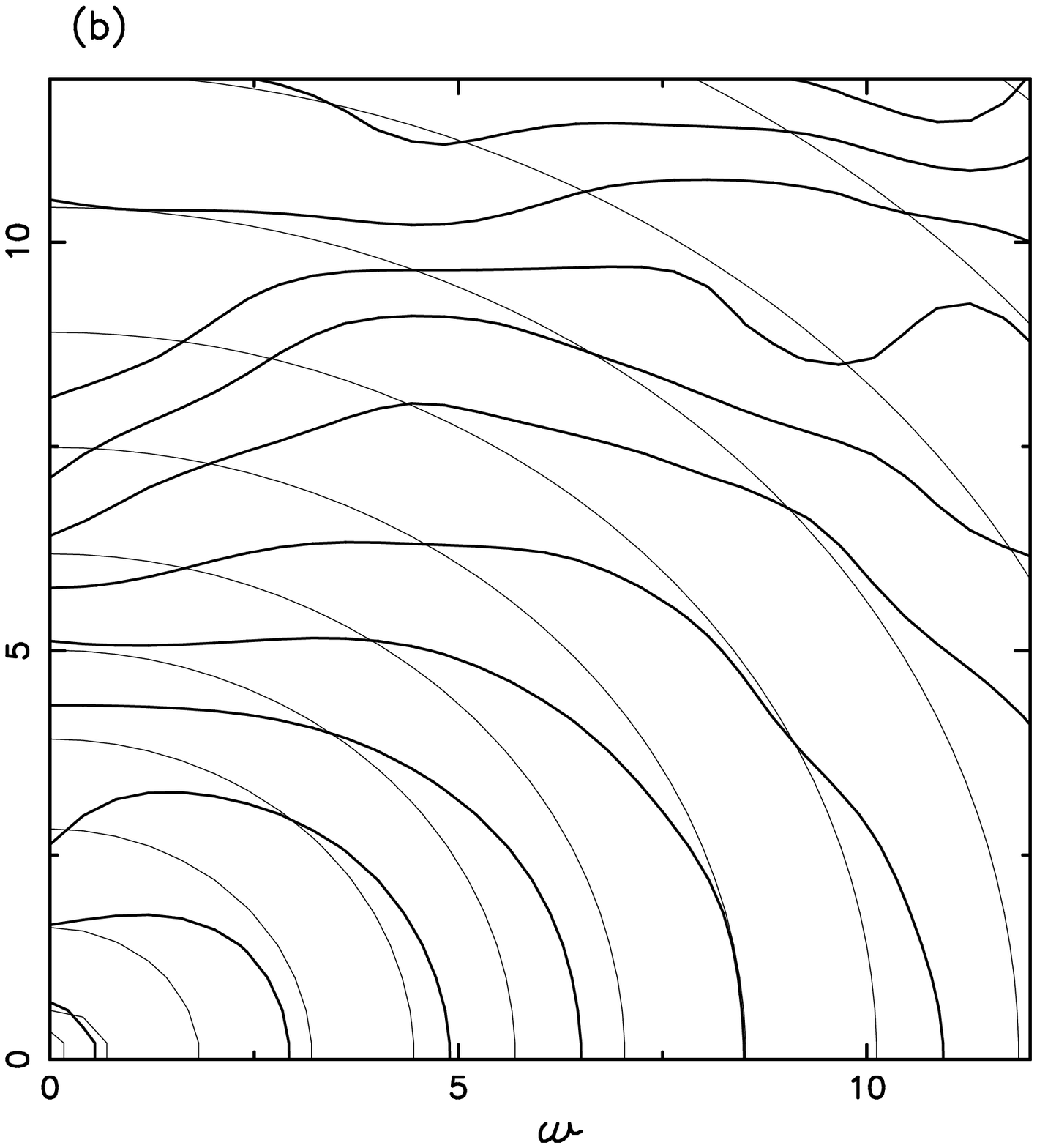}
\caption{ }
\end{figure}

\begin{figure}
\plotone{figure11.ps}
\caption{ }
\end{figure}

\begin{figure}
\plotone{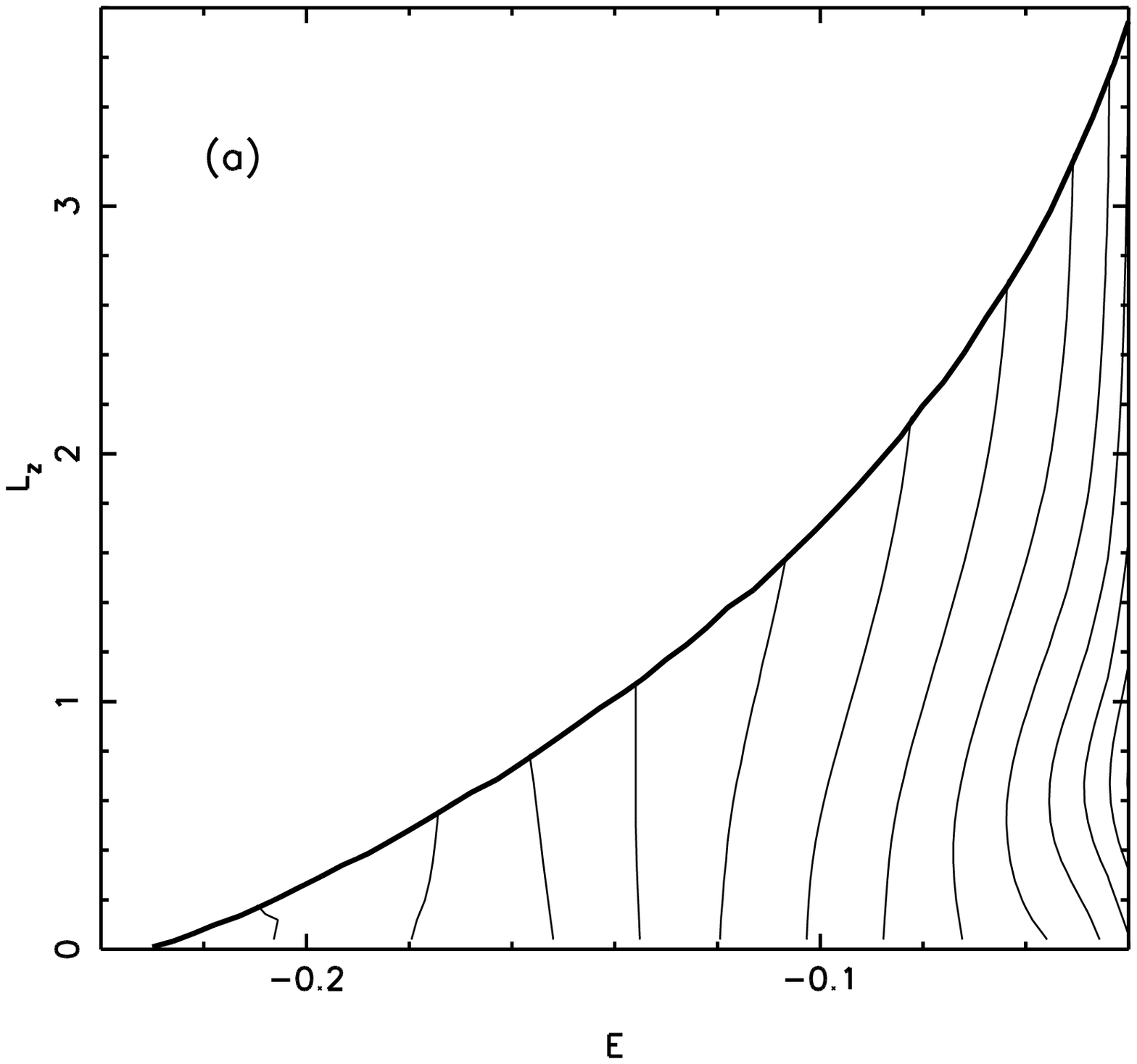}
\caption{ }
\end{figure}

\setcounter{figure}{11}

\begin{figure}
\plotone{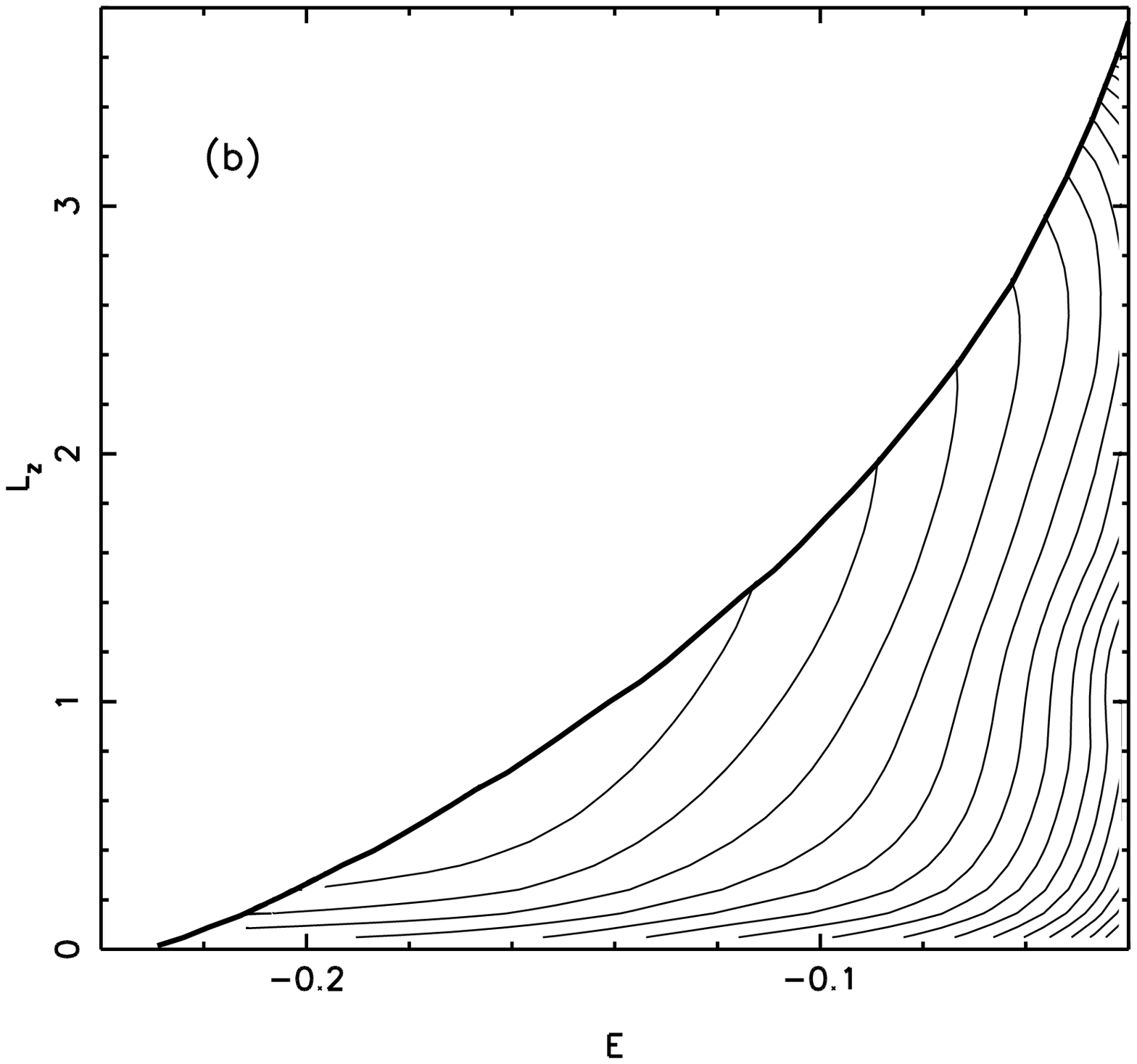}
\caption{ }
\end{figure}

\end{document}